\documentclass[twocolumn,tighten,twocolappendix]{aastex631}

\usepackage[T1]{fontenc}
\usepackage{hyperref}
\usepackage{lmodern}
\usepackage{amsmath,amssymb}
\usepackage[dvipsnames]{xcolor}
\usepackage{amsmath,amssymb}
\usepackage[dvipsnames]{xcolor}
\usepackage{graphicx}
\usepackage{placeins}
\usepackage{multirow}

\newcommand{\Msun}{M_\odot}
\newcommand{\Rsun}{R_\odot}
\newcommand{\MBH}{M_{\mathrm{BH}}}

\newcommand{\Ms}{M_\star}

\newcommand{\Rs}{R_\star}

\newcommand{\rg}{r_{\rm g}}

\newcommand{\torb}{t_{\rm orb}}
\newcommand{\ergs}{\mathrm{erg\,s^{-1}}}
\newcommand{\lta}{\lower 2pt \hbox{$\, \buildrel {\scriptstyle <}\over {\scriptstyle \sim}\,$}}
\newcommand{\gta}{\lower 2pt \hbox{$\, \buildrel {\scriptstyle >}\over {\scriptstyle \sim}\,$}}

\newcommand{\ulemsout}{\leavevmode \bgroup \ULdepth=-.55ex \ULset}
\makeatletter
\newcommand{\editsout@strike@citep}[1]{\mbox{\editsout@orig@citep{#1}}}
\newcommand{\editsout@strike@citet}[1]{\mbox{\editsout@orig@citet{#1}}}
\newcommand{\editsout@strike@cite}[1]{\mbox{\editsout@orig@cite{#1}}}
\newcommand{\editsout}[1]{%
  \def\editsouttemp{#1}%
  \let\editsout@save@citep\citep
  \let\editsout@save@citet\citet
  \let\editsout@save@cite\cite
  \let\citep\editsout@strike@citep
  \let\citet\editsout@strike@citet
  \let\cite\editsout@strike@cite
  \expandafter\ulemsout\expandafter{\editsouttemp}%
  \let\citep\editsout@save@citep
  \let\citet\editsout@save@citet
  \let\cite\editsout@save@cite
}
\AtBeginDocument{%
  \global\let\editsout@orig@citep\citep
  \global\let\editsout@orig@citet\citet
  \global\let\editsout@orig@cite\cite
}
\makeatother

\begin{document}
\title{Tidal Disruption Events in Active Galactic Nuclei: Thermal Radiation and Jets}

\author{Nimrod Tripto}
\affiliation{Racah Institute of Physics, The Hebrew University of Jerusalem, Israel}
\email{nimrod.tripto@mail.huji.ac.il}

\author{Tsvi Piran}
\affiliation{Racah Institute of Physics, The Hebrew University of Jerusalem, Israel}

\author{Julian Krolik}
\affiliation{Department of Physics and Astronomy, Johns Hopkins University, Baltimore, USA}

\begin{abstract}
Over the past decade, dozens of Tidal Disruption Events (TDEs) have been identified, with many more expected from upcoming surveys. 
A small subset has been observed to launch transient, highly luminous  relativistic jets,
for which the origin of the requisite magnetic flux has not been identified.
A non-negligible fraction of TDEs are expected in Active Galactic Nuclei (AGNs), where a magnetized accretion flow is already present. 
As shown first by \citet{Chan2019}, the collision of a TDE debris stream with this flow can substantially alter both disk and debris dynamics, leading to a possibly Eddington-limited flare.
Here, we use the Athena++ radiation magnetohydrodynamic (RMHD) code to simulate the time near peak fallback rate of a TDE in an AGN, studying the debris-disk interaction, the radiation emitted, and the accumulation of magnetic flux near the horizon.
In the example simulated, the accretion rate climbs to $\sim 12\,\dot{M}_{\rm Edd}$, 
while the escaping solid angle-integrated luminosity approaches $\sim L_{\rm Edd}$.
The photon luminosity is strongly asymmetric, brighter in the hemisphere from which the debris strikes the disk and dimmer in the hemisphere toward which the debris travels.  The photon spectrum peaks in the EUV.
The polar magnetic flux grows $\sim 33$-fold, enough to raise the Blandford--Znajek luminosity by 
a factor $\sim 10^3$, reaching a level sufficient to power observed TDE jets.
The polar magnetic flux is also asymmetric, likely making any jet similarly asymmetric, but in the opposite sense to the photon flux.

\end{abstract}

\section{Introduction}

Tidal disruption events (TDEs) occur when a star passes close enough to a massive black hole to be torn apart by tidal forces. Roughly half the stellar debris becomes unbound, while the other half remains bound and returns to the black hole on highly eccentric orbits \citep{Rees1988,Phinney1989}. The returning debris produces a luminous flare lasting from months to years \citep[see e.g.][]{Gezari2021}.
Most observed TDEs appear to be dominated by thermal optical/UV \citep{Gezari2021,Guolo+2024}, but a small subset show luminous hard X-ray emission, rapid variability, and other signatures associated with relativistic jets \citep{Bloom2011,Burrows+2011,Cenko+2012,Brown+2015,Andreoni+2022,Pasham+2023}.

Most TDE studies consider an isolated black hole, but roughly $\sim10\%$ of galaxies host an AGN \citep[e.g.,][]{Ho2008,Kauffmann2003,Seth2008,Foord2017}. At least this fraction of TDEs is therefore expected to occur in AGNs.  A number of mechanisms have been suggested that may enhance the rate per AGN \citep{Karas2007,KaurStone2025,Just2012,Kennedy2016,Wang2024,Panamarev2018,MWang2024}, increasing the overall fraction of such events relative to the total number of TDEs.

When a TDE takes place within an AGN, the disruption phase is more or less unchanged as the AGN's disk is extremely dilute in comparison to a star. 
The situation changes when the returning debris stream collides with the pre-existing accretion disk, shocking the gas and driving rapid inflow in the region inside the debris-disk impact point \citep{Chan2019}. The collision drastically changes both the debris stream and the disk dynamics, as well as the radiative output and how magnetic flux is carried toward the black hole. Following these effects requires radiation-MHD simulations in which the disk structure, magnetic stresses, shocks, and radiative losses evolve together.

A small but growing number of TDE candidates have recently been identified within AGN 
\citep{Blanchard2017, Mattila+2018, Sun2025, Dgany2026} exhibiting behavior not well explained by disruptions around isolated black holes.
It has been speculated that the Bowen fluorescence lines seen in some TDEs are powered by the AGN host
\citep{Trakhtenbrot2019}, 
but this is an ambiguous signal because ordinary TDEs may also produce them
\citep{Krolik+2025}. This uncertainty further underscores the need for a theoretical framework for such TDEs, tailored to the AGN environment.

\citet{Chan2019} carried out the first study of a TDE debris stream colliding with an AGN disk. Their hydrodynamic simulations
considered an idealized Shakura-Sunyaev disk. They showed that within a fraction of the debris return time, the stream-disk collision excites shocks, drives rapid inflow, and depletes the disk interior to the impact radius. In these simulations, the energy dissipation and mass inflow rates often exceeded those of the unperturbed disk by orders of magnitude, although much of the dissipated energy was advected into the black hole without escaping as radiation.  An Eddington-scale luminosity is expected, as in any case in which an optically thick 
source is powered by gravitation \citep{Krolik+2025}. 
Those runs were purely hydrodynamic, so they could not follow radiative transport, magnetic stresses, or magnetic flux accumulation.

Subsequent hydrodynamic studies extended this picture. \citet{Ryu2024} simulated in-plane, prograde, and retrograde disruptions in an AGN disk and found that continuous stream-disk interaction can fully mix the debris into the disk before it returns coherently. \citet{Zhang2025} explored the dependence on the star's orbital inclination and included Schwarzschild apsidal precession, identifying a two-phase light curve. Both studies, like \citet{Chan2019}, adopt a thin, gas-pressure-dominated disk with adiabatic or locally approximated cooling, and neither evolves the radiation field or includes dynamically important magnetic pressure.

\citet{Tripto2026} gave analytic estimates for magnetic-flux transport driven by the stream-disk collision. 
Assuming that the local flux density was of order the local field intensity, they argued that
for a wide range of parameters, and under conservative assumptions, the sharply-increased mass accretion when a TDE occurs in an AGN could supply enough horizon-threading flux to power a transient jet.

Here we report results derived from a radiation-MHD simulation 
of the stream-disk interaction at pericenter. Our simulation 
simultaneously gives a good description of both the radiation transfer and the transport of magnetic flux toward the black hole.
It is distinguished from previous work by using an initial structure drawn from a recent radiation-MHD simulation of a stationary accretion disk \citep{Jiang2019}.
This structure differs substantially from the Shakura-Sunyaev models considered in \citet{Chan2019,Chan2020,Chan2021}, or the semi-analytic thin AGN disk solutions used in \citet{Ryu2024, Zhang2025}.

\section{Methods}

\subsection{Athena++ with radiation and magnetic fields}

We carry out 3D radiation magnetohydrodynamic (RMHD) simulations using Athena++ \citep{Athena++2019}. The code evolves the equations of ideal magnetohydrodynamics together with a time-dependent 3D radiation transport module. 
The simulations are performed in spherical-polar coordinates $(r,\theta,\phi)$ centered on the black hole. We use the pseudo-Newtonian \citet{PaczynskiWiita1980} gravitational potential,
\begin{equation}
\Phi_{\rm PW} =
-\frac{GM_{\rm BH}}{r-2\rg},
\end{equation}
where $\rg\equiv GM_{\rm BH}/c^2$. This approximation reproduces the main relativistic effect relevant for the inner disk --- the existence of an innermost stable circular orbit --- while keeping the calculation Newtonian.

Magnetic fields are evolved with the constrained-transport algorithm in Athena++, which preserves $\nabla\cdot{\bf B}=0$ to numerical precision. That property matters when we ask whether the debris impact changes the magnetic flux near the black hole. Magnetic pressure and Maxwell stresses also enter the disk structure and angular-momentum transport.

The radiation field is described by a large number of angle-groups evolved simultaneously with the gas, including energy and momentum exchange between radiation and matter.  The transport equations are ``gray'', i.e., both opacity and emissivity are averaged over photon frequency assuming a blackbody spectrum.
Treating radiation transfer explicitly is essential because radiation affects the subsequent evolution through cooling, radiation pressure forces are comparable to gravity, and radiation viscosity is significant \citep{Jiang2019}.

\subsection{Specifics}
\subsubsection{Initial AGN disk}

The initial state of the accretion flow is taken from a late-time ($t \simeq 4 \cdot 10^4 \MBH$) snapshot of the ``AGN0.07'' simulation of \citet{Jiang2019}\footnote{The code unit of length is $\rg \equiv GM_{\rm BH}/c^2$; taking $G=c=1$, both distance and time are in units of $M_{\rm BH}$.}.
 This simulation followed a sub-Eddington radiation-MHD accretion disk around a supermassive black hole ($\MBH =5 \cdot 10^8\,\Msun$), with a time-averaged accretion rate $\approx 0.08\dot{M}_{\rm Edd}$ 
 during roughly the second half of the simulation\footnote{\citet{Jiang2019} defines $L_{\rm Edd}$ using $\kappa=0.34 \, \text{cm}^2\,\text{g}^{-1}$, while we use $\kappa=0.4 \, \text{cm}^2\,\text{g}^{-1}$. Throughout the paper, we implicitly convert quoted values, which is why their ``AGN0.07'' run has $\dot{M}/\dot{M}_{\rm Edd} \sim 0.08$.}. Throughout this paper, we define
 $\dot{M}_{\rm Edd}\equiv 10\,L_{\rm Edd}/c^2$ with a fiducial radiative efficiency $\eta=0.1$.
The initial magnetic field was axisymmetric, with a pair of nested quadrupolar field loops above and below the disk's midplane. 

In the initial state of the \citet{Jiang2019} simulation, a thick torus at $40-100 \rg$ feeds a central region whose accretion flow is determined dynamically. Regrettably, this simulation ran long enough to reach inflow equilibrium only inside $\simeq 15\, \rg$, a distance comparable to our impact radius. Dynamics driven by the stream collision  cannot be completely separated from the continued relaxation of the $r>15\,\rg$ region. However, the abrupt changes we observe occur on timescales much shorter than the undisturbed disk's inflow timescale, and can therefore be attributed to the stream collision.

The principal limitation created by the lack of inflow equilibrium outside $\simeq 15 \rg$ is that the mass stored in the initial torus ($\simeq 40 - 100 \rg$) may contribute excess mass to the disk's atmosphere at larger radii. This is unlikely to significantly influence our results, for two main reasons. First, as shown in Sec.~\ref{sec:mass_sources}, the late-time mass inside $30\rg$ is split roughly equally between debris and gas that began outside $30\rg$; thus, even if mass inflow from outside $30\rg$ is overestimated, its 
contribution remains constrained. Second, our simulation is not run for a long duration, so the slower effects of matter inflowing from further-out regions remain limited. However, the existence of this torus prevents us from exploring the longer-term interaction of the debris stream with outer regions of the disk \citep{Chan2021}.

We rescaled this system to be consistent with 
$M_{\rm BH}=5\cdot 10^6\,\Msun$ while maintaining $\dot M \approx 0.07 {\dot M}_{\rm Edd}$. This was necessary because $\MBH \sim 5\cdot10^{8}\,\Msun$ is excluded for TDEs involving main-sequence stars by the Hills criterion \citep{Hills1975}.
The conversion is described in Appendix~\ref{app:unitconverstion}. The rescaled disk was evolved for several orbits before the debris stream was introduced to reduce transients and assess its stability. 

The disk structure computed by \citet{Jiang2019} is quite different from what an $\alpha$-model would predict, beginning with the fact that the ratio of stress to total pressure is $\propto r^{-1}$, rather than constant. Moreover, the AGN0.07 disk is magnetically supported and has a surface density (and therefore optical depth) a factor $\simeq 10$ below that of a thin $\alpha$-disk with the same accretion rate. 
At the impact radius, $r\simeq 15\,\rg$, the disk prior to stream injection has midplane density $\rho_{\rm d}\simeq 10^{-9}\,\mathrm{g\,cm^{-3}}$, surface density $\Sigma_{\rm d}\simeq 4\cdot 10^{2}\,\mathrm{g\,cm^{-2}}$, and Rosseland-mean optical depth $\tau_{\rm R}\simeq 100$. The photosphere lies only $\simeq 6^\circ$ from the midplane. Very much unlike an $\alpha$-disk, the gas pressure is smaller than both the magnetic and radiation pressures by more than a factor of $10^{3}$.

\subsubsection{Debris stream}
\label{sec:debris_stream}

\begin{figure}[!t]
  \centering
  \includegraphics[width=\linewidth]{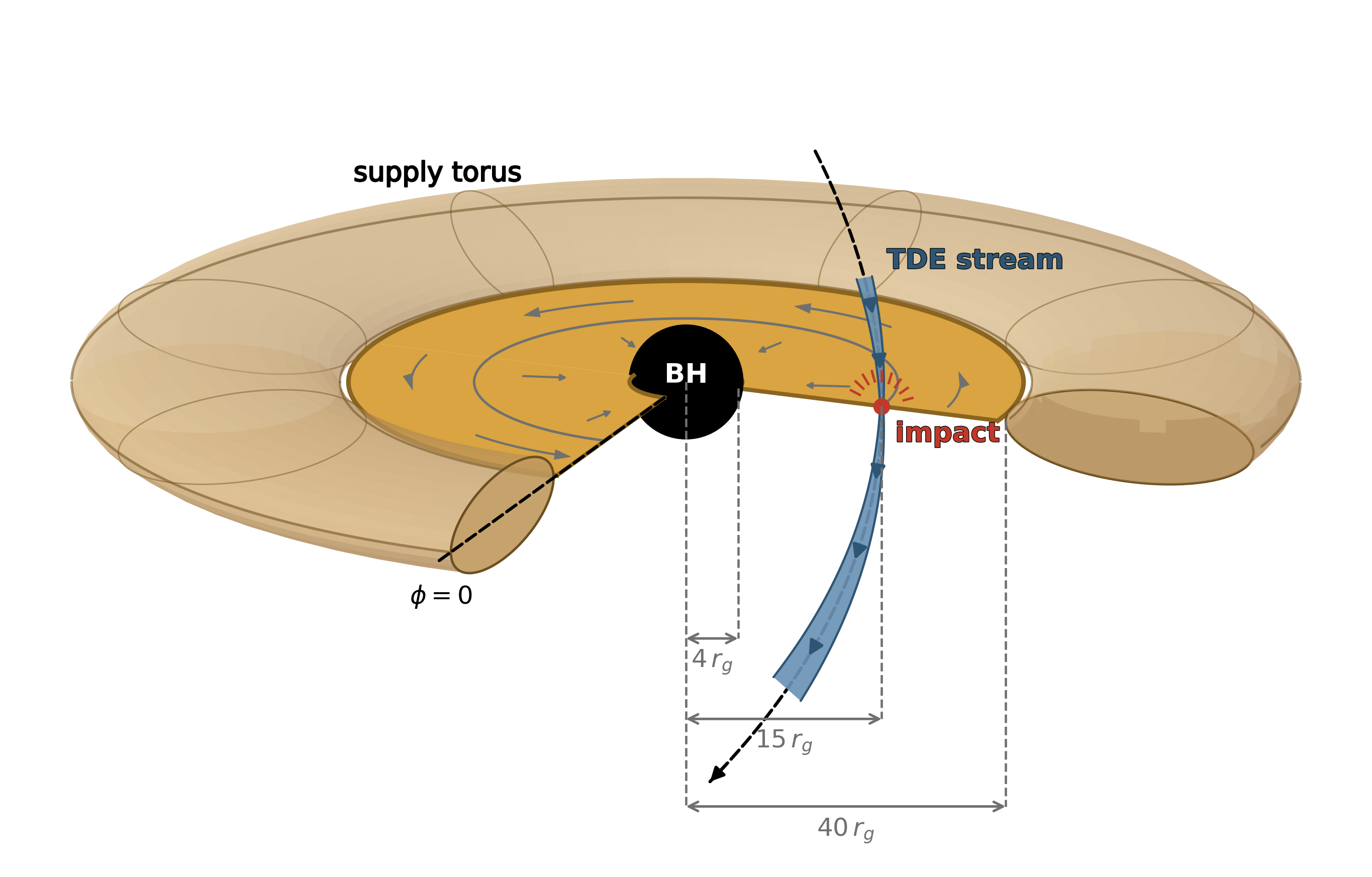}
  \caption{A schematic picture of the simulated scenario. The pre-existing AGN disk is described following \citet{Jiang2019} as a $40\rg$ disk supported by radiation and magnetic pressure. This disk is fed by a large mass reservoir, a supply torus centered at $80\rg$ with an inner edge of $40\rg$. The disk luminosity is 0.08 $L_{\rm Edd}$.
  A TDE debris stream (blue arrows: injection direction) is injected along a nearly parabolic trajectory into this disk and impacts it at $r_{\rm imp}=15\,\rg$ and $\phi=1.52$. 
  Because $\dot{M}_{\rm s}/\dot{M}_{\rm d}\sim400$ (Sec.~\ref{sec:comparison_with_chan}), the stream punctures the thin disk. 
  A small fraction of its mass is left behind in the disk. Disk matter is pushed downward, first tilting the disk and then spreading it vertically  (Fig.~\ref{fig:rho_slices}).}
  \label{fig:schematic}
\end{figure}

Our injection scheme follows the same general approach as that of \citet{Chan2019}, as depicted in Fig.~\ref{fig:schematic}. We inject the stream  $\sim4 \, \rg$ above the equatorial plane, so that it crosses the plane at $r=15\,\rg$. For a star with $\Ms\approx 0.8\,\Msun$ this is approximately the physical tidal radius \citep{Ryu+2020d}. It is injected with a velocity corresponding to an almost parabolic trajectory (App.~\ref{app:stream}). The orbit is nearly perpendicular to the disk, with inclination $i \simeq 86^\circ$. The injected mass current is $\dot{M}_{\rm s}=8.3\cdot 10^{25}\,\mathrm{g\,s^{-1}}=12\,\dot{M}_{\rm Edd}$, which is $\sim 170$ times the unperturbed disk accretion rate of $0.07\,\dot{M}_{\rm Edd}$.
The initial inner disk mass is $\sim2.5\cdot 10^{-4} \, \Msun$ within $r<30\,\rg$ and $\sim 4.5 \cdot 10^{-5} \, \Msun$ within $ r<15\,\rg$.  Thus, the stream injects comparable masses in just $\sim0.7\torb$ and $\sim \torb/8$, respectively, where $\torb\approx 365\,G\MBH/c^{3} {\sim 0.1 \, {\rm d}}$ is the Keplerian orbital period at $15\,\rg$. 
The stream has a transverse width of $\Rsun\simeq 0.1\,\rg$ by $ 3\Rsun\simeq 0.3\,\rg$ at injection, i.e.\ $\sim 1.2$--$4\Rs$.

Throughout the simulation $\dot{M}_{\rm s}$ is held constant, 
corresponding to the peak fallback rate. This is consistent because our simulation ran for $43.5$ orbits at $15\,\rg$, or $4.4$~d, which is only $\simeq 0.06\,t_0$, where $t_0\simeq 75$~d is the debris fallback time for $M_{\rm BH}=5\cdot 10^{6}\,\Msun$ and $\Ms\approx 0.8\,\Msun$ \citep{Ryu+2020a,Ryu+2020b}.
A more detailed description of the stream injection parameters and procedure is given in App.~\ref{app:stream}.

\subsubsection{\texorpdfstring{Comparison with \citet{Chan2019}}{Comparison with Chan et al.\ (2019)}}

\label{sec:comparison_with_chan}

The controlling parameter of this simulation, as in \citet{Chan2019}, is the mass-current ratio $\dot{M}_{\rm s}/\dot{M}_{\rm d}$, where $\dot{M}_{\rm d}=\Sigma_{\rm d}R_{\rm s}\sqrt{GM_{\rm BH}/r_{\rm imp}}$ is the azimuthal mass current of the disk under the stream footprint, with $r_{\rm imp}$ being the impact radius. With $\Sigma_{\rm d}\simeq 4\cdot 10^{2}\,\mathrm{g\,cm^{-2}}$ and $R_{\rm s}\simeq \Rsun\simeq 0.1\,\rg$, one finds $\dot{M}_{\rm d}\simeq 2\cdot 10^{23}\,\mathrm{g\,s^{-1}}$, so $\dot{M}_{\rm s}/\dot{M}_{\rm d}\sim 400$. 

A Shakura \& Sunyaev disk with our black hole mass and impact radius already gives $\dot{M}_{\rm s}/\dot{M}_{\rm d}\sim 30$ \citep{Shakura1973}, a factor of a few above the maximum of $\dot{M}_{\rm s}/\dot{M}_{\rm d}\sim 4$ considered by \citet{Chan2019} even before any change in disk model. The AGN0.07 disk's surface density is $\simeq 11$ times lower than an $\alpha$-disk at the same parameters, pushing $\dot{M}_{\rm s}/\dot{M}_{\rm d}$ a further order of magnitude to $\sim 350$, consistent with the $\sim 400$ found above and well outside the range $[0.05,4]$ explored by \citet{Chan2019}.
Thus, their hydrodynamic results remain a qualitative guide, but with such a large contrast in the mass-current ratio, it is difficult to quantitatively compare between the simulations.

One of the notable results of the difference in $\dot{M}_{\rm s}/\dot{M}_{\rm d}$ is that the stream does not merely disrupt the inner disk: unlike in \citet{Chan2019}, it 
also replenishes it with a comparable or greater mass of fresh material. This point is further discussed in Sec.~\ref{sec:mass_sources}.

\subsubsection{Numerical details}

In our spherical coordinate system, the radial coordinates of cell-centers are logarithmically spaced, with $r_{\rm in}=4\rg$ and
$r_{\rm out} = 1.6\cdot 10^3 \rg$.
The grid covers the full polar and azimuthal ranges, $0\leq\theta\leq\pi$ and
$0\leq\phi<2\pi $.
The base grid has $64\times32\times64$ cells in $r,\theta,\phi$ respectively. Cell sizes are self-similar, with $\Delta r/r=\Delta\theta=\Delta\phi$ at every refinement level: they are uniform in logarithmic measure but not in physical size. The base values are $\Delta r/r=\Delta\theta=\Delta\phi\simeq 9.8\cdot 10^{-2}$ ($\Delta\theta=\pi/32$). We use static mesh refinement, following the AGN0.07 setup of \citet{Jiang2019}, to increase the resolution in the disk midplane and around the stream-disk interaction region. This midplane refinement reaches $\Delta r/r=\Delta\theta=\Delta\phi=6.1\cdot 10^{-3}$. We additionally refine the grid around the stream injection point and its trajectory with two rectangular regions in $(r,\theta,\phi)$. The first encloses the injection site with boundaries defined by
$
15\rg \leq r\leq 17\rg$,
$1.25\leq\theta\leq 1.97$,
and $1.42\leq\phi\leq 1.62.
$
The second extends the refined zone along the incoming stream, bounded by
$16\rg \leq r\leq 21\rg$,
$1.97\leq\theta\leq 2.3$, 
and $1.38\leq\phi\leq 1.62$.
In the two stream regions the finest cells reach $\Delta r/r=\Delta\theta=\Delta\phi=1.5\cdot 10^{-3}$, corresponding to $0.025\,\rg$ per cell at the impact radius. 
We define the northern hemisphere as the one from which the stream flows, and the southern hemisphere as the one into which it flows.

The radial boundary conditions are outflow conditions both at the inner and outer boundaries. The polar boundaries use the standard coordinate-singularity treatment, in which ghost zones are filled from the corresponding cells across the pole with vector components appropriately sign-flipped.
The azimuthal direction is periodic. No matter is allowed to enter the simulation domain except for the source terms that inject the debris stream.

The radiation field is discretized using the default angular quadrature of \citet{Davis2012} at fourth order, corresponding to 80 discrete ray directions per cell. The radiation boundary conditions are very similar to the hydrodynamic boundary conditions: pure outflow on the radial boundaries,
polar in the $\theta$ direction, and periodic in $\phi$. The opacity is computed locally from the gas density and temperature using tabulated opacity \citep{Zhu2021}. Atomic features (both scattering and absorption), electron scattering (with energy exchange), and free-free are all included.  The emissivity is defined in terms of the usual thermal reciprocity.

Density, pressure, and temperature floors are imposed in low-density regions to maintain numerical stability. In the fiducial run these are $\rho_{\rm floor}=10^{-14}~\mathrm{g\,cm^{-3}}$, $P_{\rm floor}=2.7\cdot 10^{-3}~\mathrm{dyn\,cm^{-2}}$, and $T_{\rm floor}=3.2\cdot 10^{3}~\mathrm{K}$. For photosphere detection, because floor-density gas alone would give $\tau \sim 1$ quickly and produce a spurious photosphere, cells with $\rho < 2\,\rho_{\rm floor}$ are assigned a negligible density, $10^{-4}\,\rho_{\rm floor}$, in these integrals.

\section{Results}

\subsection{Dynamic disk response}
\label{sec:results_overview}
\begin{figure*}[!hbt]
  \centering
    \includegraphics[width=0.9\linewidth]{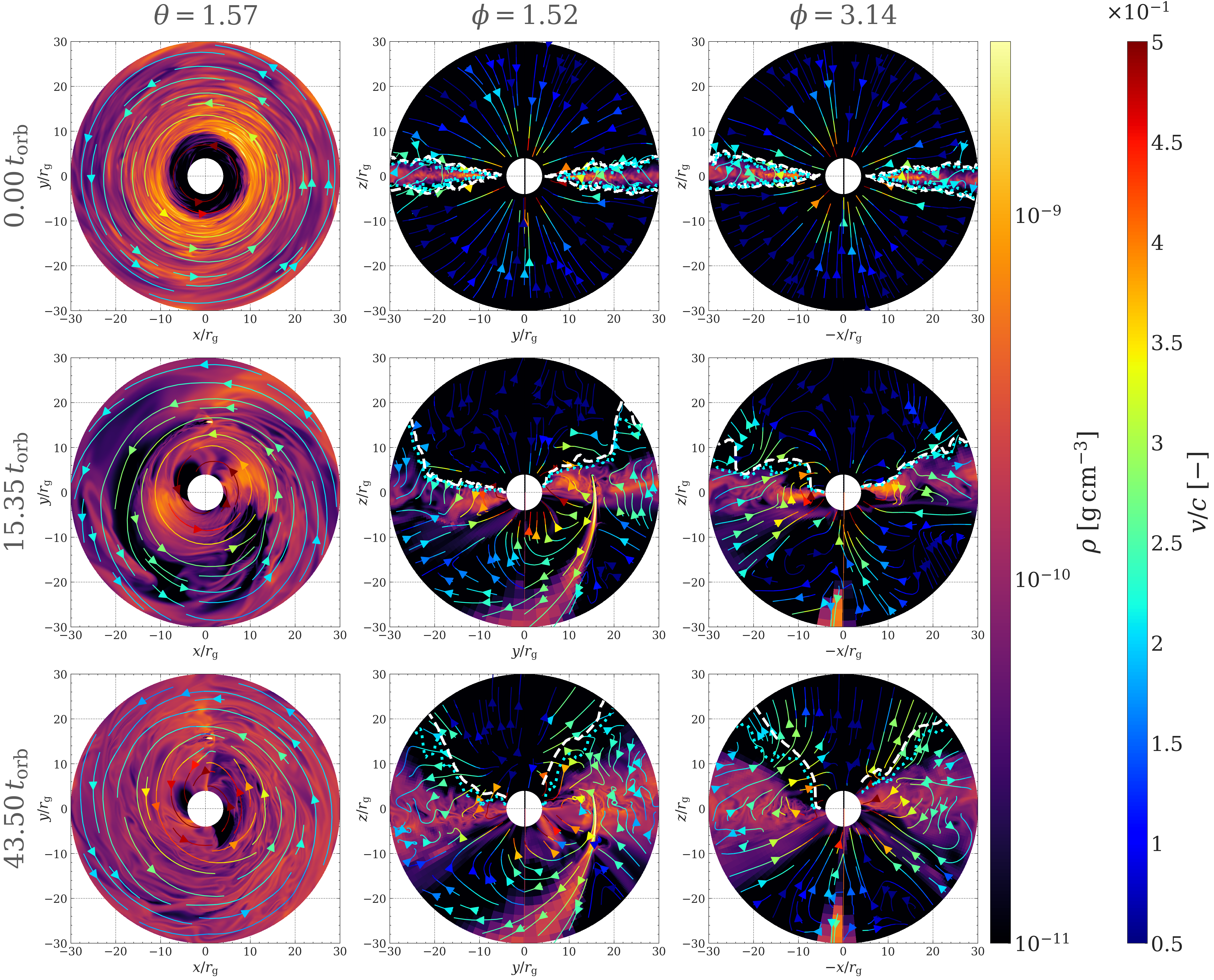}
  \caption{Three orthogonal cuts through the domain showing density slices and velocity 
  streamlines at three epochs.
The top row shows the initial disk prior to stream injection, while the middle and bottom rows show the flow after the stream-disk collision has driven strong asymmetries, spiral shocks, tilting and thickening of the disk, and partial evacuation of the inner disk. Times are labeled in units of orbits at $15\rg$. The dotted cyan curves mark the thermalization photosphere, defined by $\tau_{\rm eff}=\sqrt{\tau_{\rm abs}(\tau_{\rm abs}+\tau_{\rm sc})}=1$, and the dashed white curve marks the scattering photosphere from the Rosseland-mean optical depth; both are integrated inward from the outermost $z$ in the domain. After stream injection, the south photosphere lies below this $r=30\,\rg$ frame; both surfaces are shown on a larger scale in Fig.~\ref{fig:far_rho_slices}. 
\href{https://drive.google.com/file/d/1jkwREwSKDNHQ-B0YP99Op8HmsSsxaZxg/view}{Full video available here.}}
  \label{fig:rho_slices}
\end{figure*}   

\begin{figure*}[!hbt]
  \centering
    \includegraphics[width=0.95\linewidth]{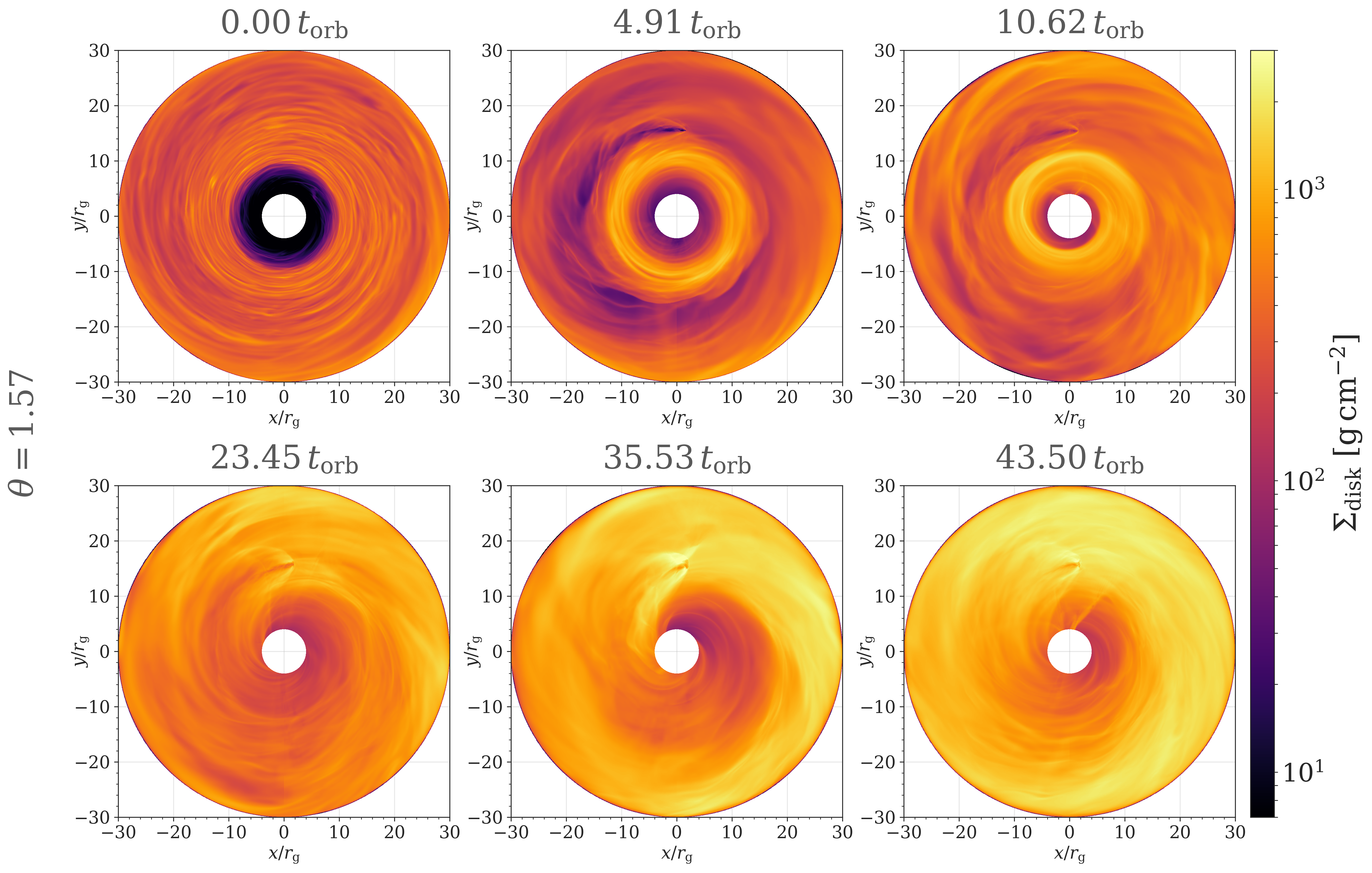}
    \includegraphics[width=0.95\linewidth]{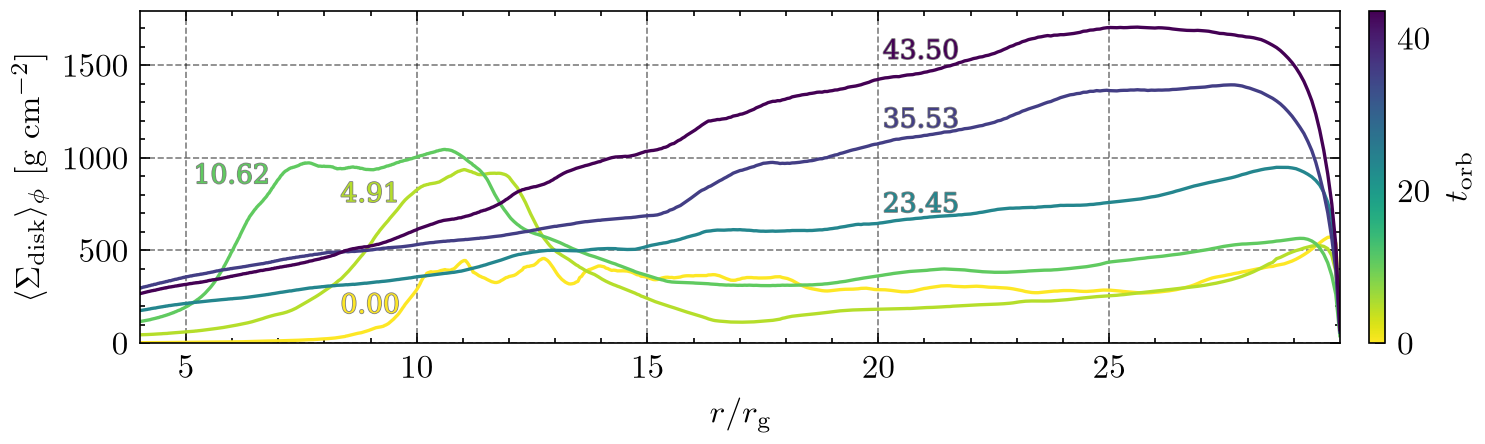}
    \caption{\textbf{(Top)} Disk surface density $\Sigma_{\rm disk} = \int\rho\,dz$ limited to $\theta\in[5\pi/16,11\pi/16]$ to include only disk matter, with outgoing stream matter removed by masking elements with $\rho > 3\cdot10^{-11} \, \text{g} \,\text{cm}^{-3}$ and either $|v_\phi|/c < 3\cdot10^{-2}$ or the combination $|v_\phi|/c < 10^{-1}$ and $|v_\theta|/c > 5\cdot10^{-2}$. At $t=0$ the disk shows low-amplitude fluctuations. 
    By $4.91$–$10.62\torb$ the stream impact opens a cavity elongated along the rotation direction while the inner disk
    density grows as matter drains inward;
    both features partially smooth out as additional matter enters the outer disk.
    By $23.45\torb$ the cavity has smoothed over and $\Sigma_{\rm disk}$ rises
    gradually outward, leaving only a faint spiral arm near $r\sim15\rg$ trailed by an under-dense lane. This lane grows more pronounced through $35.53\torb$ as the surrounding disk density continues to grow 
    ($\Sigma_{\rm disk}>10^{3}\,{\rm g\,cm^{-2}}$ outside $r\sim10\rg$), and by $43.50\torb$ the disk has relaxed to a smooth, nearly axisymmetric high-density state with only the lane persisting. The outer disk is supplied by stream material that has gained angular momentum, as shown in Fig.~\ref{fig:mass_source}. \textbf{(Bottom)} Azimuthally averaged $\langle\Sigma_{\rm disk}\rangle_\phi$ vs.\ $r/\rg$ for the same epochs. At $t=0$, the profile is empty inside the ISCO and peaks near $r\sim10$–$12\rg$. Following stream impact, the peak shifts inward and sharpens ($4.91$–$10.62\torb$) as matter piles up in the inner region, then flattens as the disk relaxes ($23.45$–$43.50\torb$); by $t=43.50\torb$ the profile rises nearly monotonically from $r\sim5\rg$ outward. \href{https://drive.google.com/file/d/1uOQOIW1F2RuL7Zw7LbYKs9o5Wy5qp7ob/view?usp=sharing}{Full video available here.}} 
  \label{fig:surface_density}
\end{figure*}

\begin{figure}[!hbt]
  \centering
    \includegraphics[width=\linewidth]{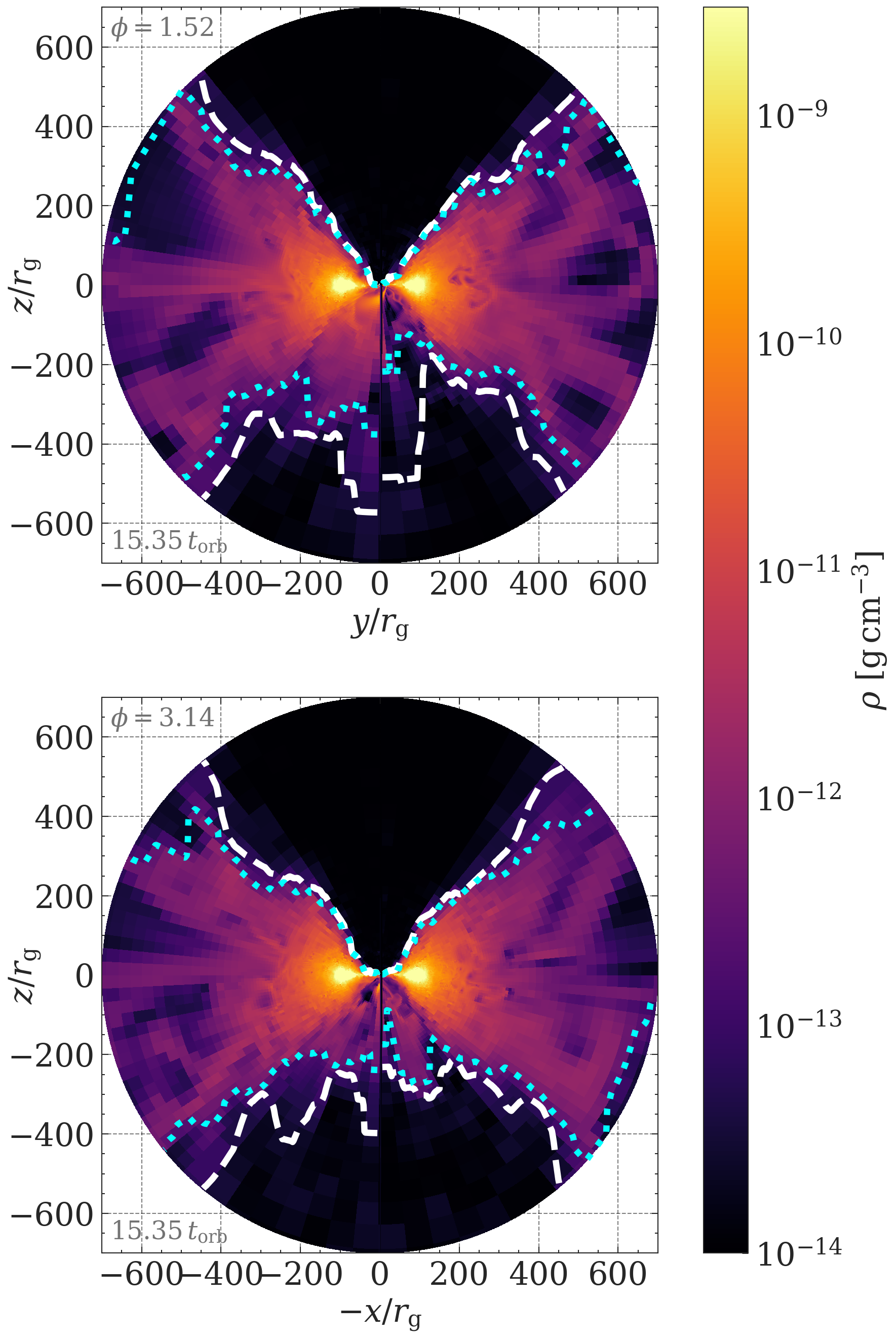}
    \caption{Meridional density slices at $t=15.35\,\torb$, the same epoch as the middle row of Fig.~\ref{fig:rho_slices}, now extending to $|x|,|z|\sim700\,\rg$ so that both photospheres are visible. Top: $\phi=1.52$; bottom: $\phi=3.14$. Curves as in Fig.~\ref{fig:rho_slices}: dotted cyan, thermalization photosphere ($\tau_{\rm eff}=1$); dashed white, scattering photosphere. The south photosphere follows the outgoing stream to $|z|\simeq200{-}600\,\rg$.}
  \label{fig:far_rho_slices}
\end{figure}

Figs.~\ref{fig:rho_slices} and \ref{fig:surface_density} depict density slices and stream lines, and the corresponding surface density evolution, before stream injection and at several later times. At $t=0$ the gas occupies a thin, nearly equatorial disk, densest near the midplane between roughly $6\,\rg$ and $20\,\rg$, with photospheres at $|z|\simeq 3{-}5\,\rg$. After impact, spiral arms and a dense incoming stream appear in the equatorial cut, matching the same spiral feature in the surface density. 
In the midplane annulus $6\,\rg \leq r \leq 25\,\rg$, the ratio of the azimuthally-averaged density $\rho_{\rm mid}(r,t)$ to $\rho_{\rm mid}(r,t=0)$ diminishes to $\sim 0.6$ at $t=15\,\torb$ in both the inner and outer parts of that range; by $t=43\,\torb$, the inner part is only $\sim 0.4$, while the density in the outer part has returned to $\sim 1$. Closer in, inside the ISCO ($r\lesssim 6\,\rg$), the midplane density rises by a factor $\sim 15$ by $t=15\torb$ and then stays roughly steady. The meridional slices show where midplane gas goes: it is driven off the equator, thickening the flow and filling high latitudes. The surface density over the same radial range does not drop correspondingly, and exceeds its pre-impact value by late times.

The stream drills through the disk, and because it cuts across the rotating gas, it carves out a gap spanning the full azimuthal extent of that radius, elongated along the direction of rotation. This resembles the largest values of $\dot{M}_{\rm s}/\dot{M}_{\rm d}$ used in \citet{Chan2019}, which is not surprising given the value we use (Sec.~\ref{sec:comparison_with_chan}). Within the inner $r=30\,\rg$ domain of Fig.~\ref{fig:rho_slices} the north photosphere rises from $|z|\simeq 3{-}5\,\rg$ to $|z|\simeq 10{-}20\,\rg$, while the south photosphere has already left that frame.
Fig.~\ref{fig:far_rho_slices} shows the same epoch on a larger scale: both photospheres are then visible, the south following the outgoing stream to $|z|\simeq200{-}600\,\rg$.

At $t=0$ the equatorial streamlines follow nearly circular orbits, while off-plane streamlines are radial, consistent with the non-rotating sparse material there. After impact, the equatorial streamlines deviate from this circular structure, manifesting the enhanced accretion rate and disturbed disk structure; near the stream impact site the streamlines point more directly inward, tracing the funneling of debris toward smaller radii. The meridional streamlines change more markedly, showing an unorganized structure in the disk, consistent with gas driven off the equator to high latitudes.

The morphology matches the hydrodynamic picture of \citet{Chan2019} in several respects: a bow-like disturbance at the stream-disk contact, spiral shocks in the inner flow, and inflow concentrated in non-axisymmetric lanes. In their runs, however, the disk interior to the impact radius thinned by roughly an order of magnitude in surface density after the impact. Here, the midplane density in $6{-}25\,\rg$ falls by only a factor of a few, while gas density rises near the black hole (factor $\sim 15$ inside the ISCO), gas is pushed to high latitudes, and undergoes continuous infall into the black hole. The surface density itself increases monotonically over most radii after an initial cavity-driven dip, reaching $\gtrsim10^3\,{\rm g\,cm^{-2}}$ by late times.

\begin{figure}[!hbt]
    \centering
    \includegraphics[width=0.45\textwidth]{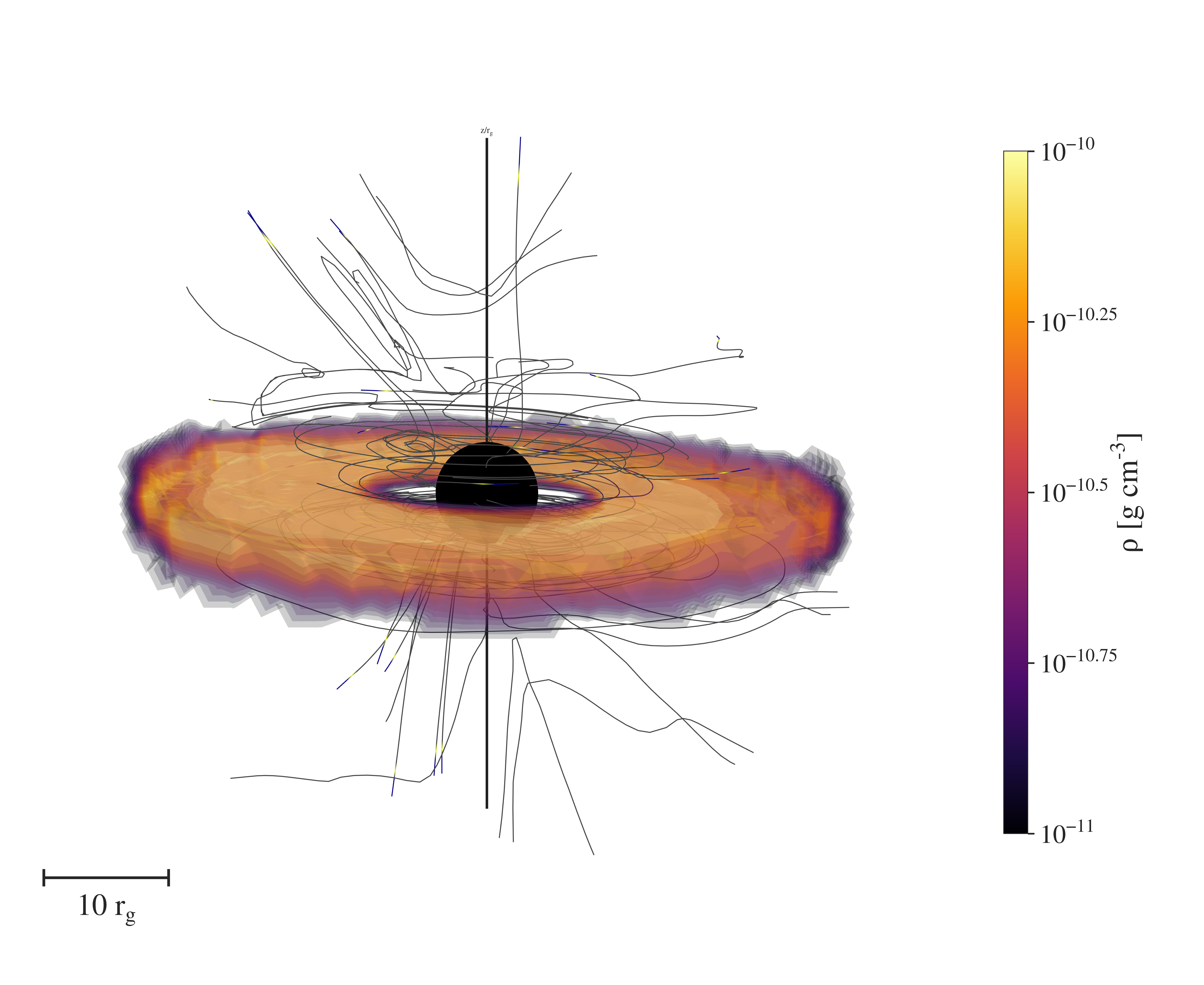}
    \includegraphics[width=0.45\textwidth]{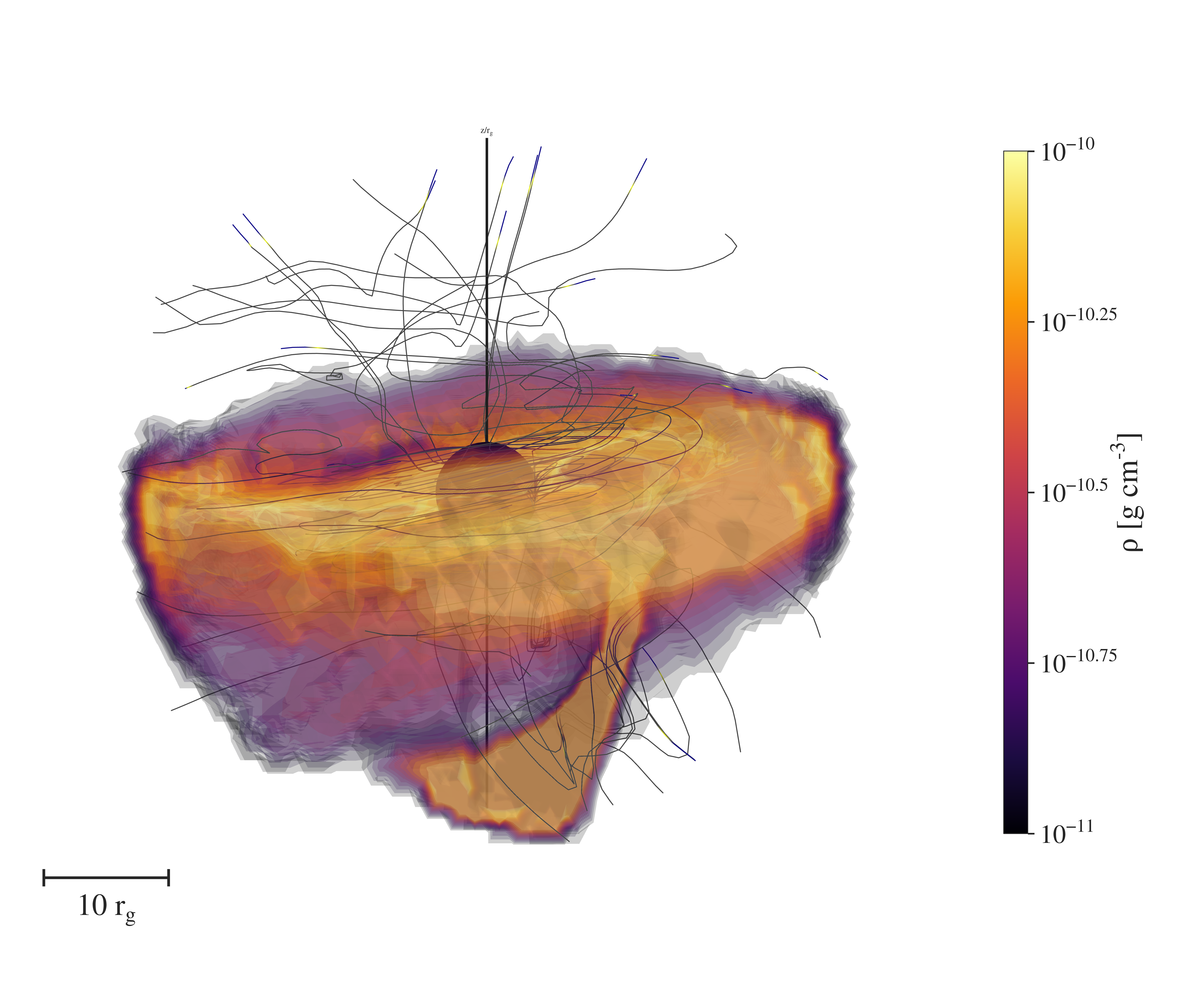}
    \includegraphics[width=0.45\textwidth]{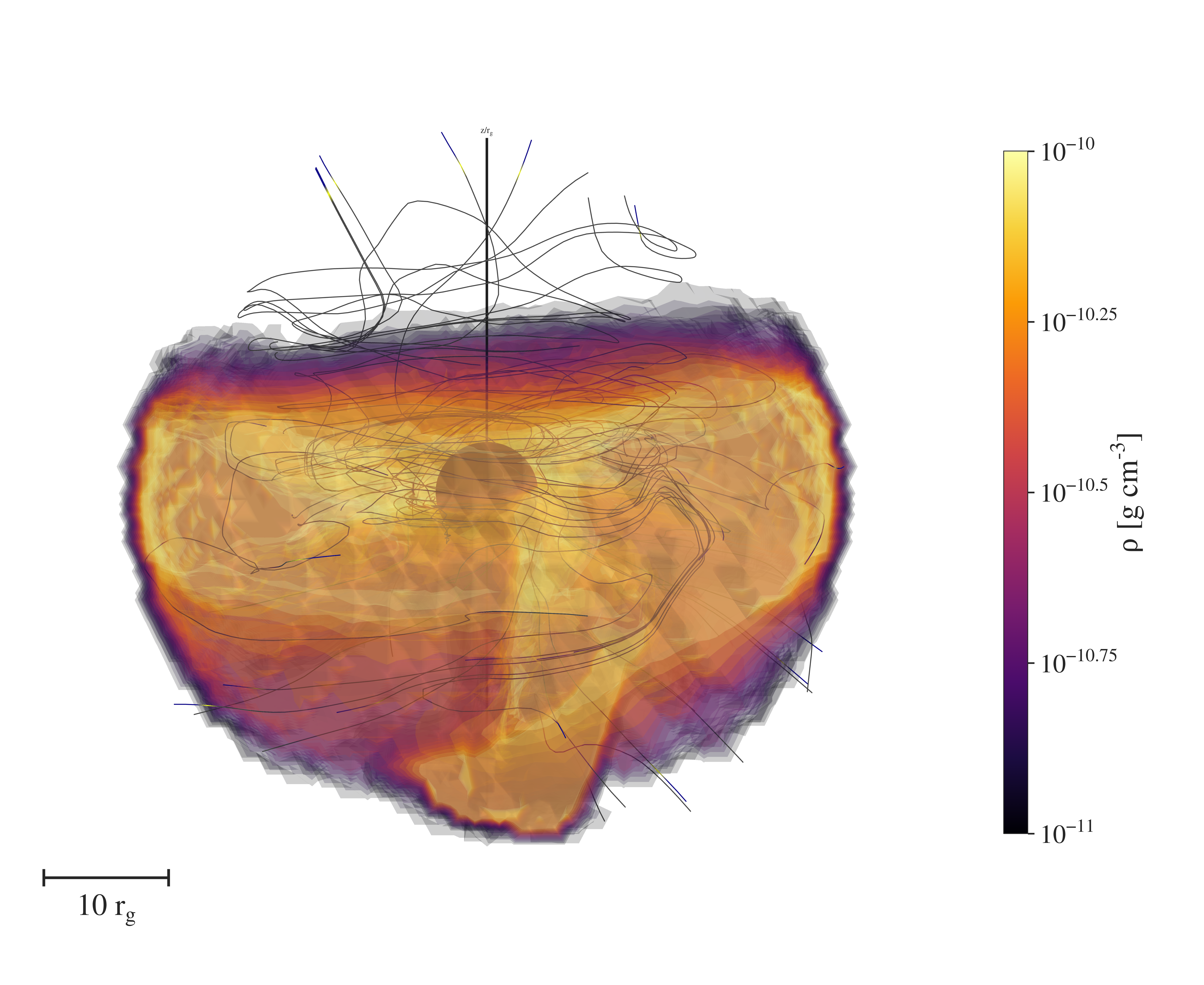}
    \caption{Three-dimensional volume renderings of $\rho$ (color) of the inner $30\rg$, with magnetic-field line tracers (curves), at the same epochs as Fig.~\ref{fig:rho_slices}. The initial thin disk becomes a thick, lopsided, vertically extended body; dense stream material remains visible as a coherent plume, especially below the midplane.}
    \label{fig:rho_3d}
\end{figure}

In 3D renderings (Fig.~\ref{fig:rho_3d}), the initial thin disk becomes a thick, lopsided, vertically extended body. Dense stream material remains visible as a coherent plume, especially below the midplane, while the former equatorial layer is distorted and inflated.

\subsubsection{Accretion rate and enclosed mass}

\label{sec:results_mdot_mass}

\begin{figure*}[!hbt]
    \centering
    \includegraphics[width=0.9\linewidth]{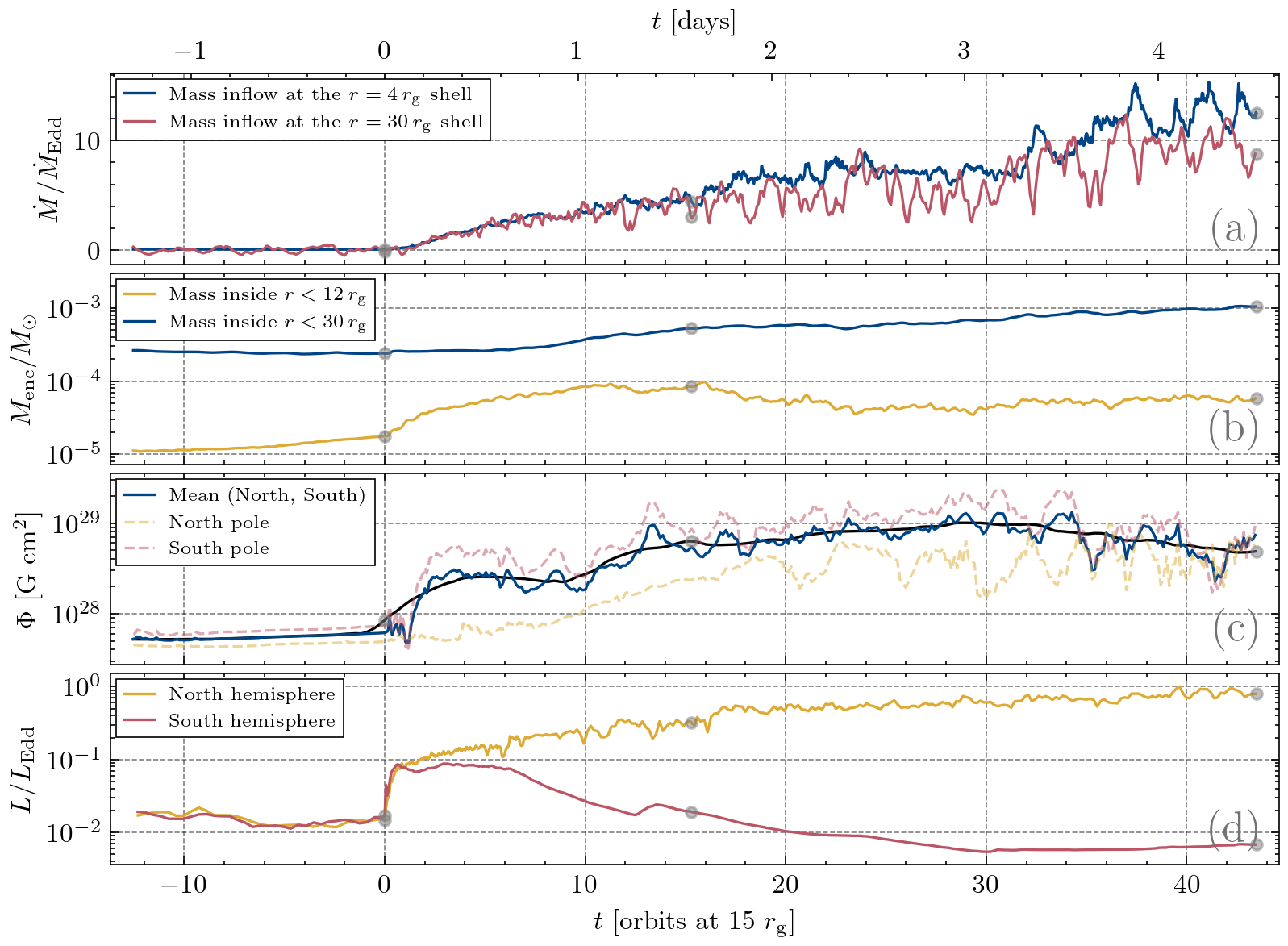}
    \caption{The dynamical response of the disk as a function of time (bottom axis in orbits at $15\,\rg$;  top axis in days). 
    Grey markers throughout indicate the epochs shown in other figures.
    \textbf{(a)} $\dot M$ in Eddington units at $r=4\,\rg$ and $r=30\,\rg$ (the unperturbed disk has $\dot{M}=0.07 \dot{M}_{\rm Edd}$). The stream-disk interaction drives a sustained rise of $\dot{M}$ to $\simeq 10$–$13\,\dot{M}_{\rm Edd}$ at both radii.
    \textbf{(b)} Enclosed disk mass within $r<12\,\rg$ and $r<30\,\rg$.  Inside both these radii, $M_{\rm enc}$ eventually increases by a factor $\simeq 5$. While the inner mass reaches a plateau, the outer mass continues to increase. 
    \textbf{(c)} North ($\theta\in[0,\pi/4]$, yellow dashed) and south ($\theta\in[3\pi/4,\pi]$, pink dashed) polar-cap magnetic flux, $\Phi = r^2\int_0^{2\pi} d\phi\, \int_{\theta_1}^{\theta_2} B_r \,\sin \theta d\theta$, before and after stream injection. The mean flux (blue) and its smoothed trend (black) are shown. Pre-impact $\Phi\simeq 6\cdot10^{27}\,\mathrm{G\,cm^{2}}$ is flat; after impact the mean rises by a factor $\simeq 4$ within $1$–$3$ orbits, and the south reaches $\simeq 2\cdot 10^{29}\,\mathrm{G\,cm^{2}}$ by $\sim 3$~d, exceeding the north by a factor $\sim 3$.
    \textbf{(d)} North (yellow) and south (pink) bolometric luminosities, obtained by integrating the vertical flux $F_z$ over the sky-plane footprint of the scattering photosphere, as seen along the $z$ line of sight (Figs.~\ref{fig:rho_slices} and~\ref{fig:far_rho_slices}).
    After impact, the thermally photon-bright north side rises toward $\sim L_{\rm Edd}$; the south declines after an initial bump because the outgoing stream obscures photons and carries them with it.}
    \label{fig:sim_time_series}
\end{figure*}

The mass accretion rate onto the inner cutout (Fig.~\ref{fig:sim_time_series}, panel (a)) rises from the unperturbed value $0.07\,\dot{M}_{\rm Edd}$ to $\simeq 10$--$13\,\dot{M}_{\rm Edd}$ over $43.5$ orbits at $15\,\rg$ ($4.4$~d), comparable to the injected stream mass current $\dot{M}_{\rm s}=12\,\dot{M}_{\rm Edd}$. 
The corresponding bolometric luminosity is shown in the same figure in panel (d) and remains near Eddington (Section~\ref{sec:results_lightcurve}). 
Relative to the unperturbed disk, $\dot{M}$ and $L$ separate by at least an order of magnitude in Eddington units, the photon trapping-dominated behavior anticipated by \citet{Chan2019}.

To quantify how the inner disk is depleted and resupplied, Fig.~\ref{fig:sim_time_series} (panel (b)) plots the enclosed disk mass within $r<12\,\rg$ as a function of time. We choose $12\,\rg$ since it is the outermost radius in the inner disk confidently free of outflowing stream matter.
The enclosed mass rises by a factor $\sim 10$, from $\sim 10^{-5}\,\Msun$ to $\sim 10^{-4}\,\Msun$ over the first $\sim 15$ orbits. It then falls to $\sim 5 \cdot 10^{-5} \Msun$ and subsequently fluctuates around that value. A midplane cavity opens (Fig.~\ref{fig:surface_density}), 
but the enclosed mass within a sphere of radius $30\rg$ does not decrease, because much of the gas is pushed out of the midplane (Figures~\ref{fig:rho_slices} and \ref{fig:rho_thetaphi}) and the inner disk is continuously resupplied from the stream and mass at larger radii. 

Fig.~\ref{fig:sim_time_series} panel (a) shows us that more mass is accreted by the black hole than is brought from large radii, and stream mass is the only other source of mass in the simulation. At late times ($t \gtrsim 35\,t_{\rm orb}$), stream material supplies $\simeq 8\,\dot M_{\rm Edd}$ of the $\simeq 12\,\dot M_{\rm Edd}$ accreted through $r = 4\,\rg$, i.e.\ about two-thirds of the injected $\dot M_s$; the remainder is pre-existing disk material. Thus, although the stream punctures the disk and part of it continues past the impact point, most of it is eventually accreted. 
Little of it falls in promptly: the bulk is deflected at the collision, swings past pericenter, and falls inward only afterward. 
This is dramatically different from ordinary TDEs, where most of the stream mass continues to orbit at $\sim 10^{3} \rg$ for at least several $t_0$ \citep{Krolik+2025}. We discuss this further in the following Sec.~\ref{sec:mass_sources}.

These results differ from the hydrodynamic picture of \citet{Chan2019}, who found a surface-density depletion of roughly an order of magnitude interior to the impact radius. The difference is most plausibly attributable to a small fraction of the stream mass joining the disk during the collision, as discussed above: because our $\dot{M}_{\rm s}/\dot{M}_{\rm d}$ is much larger than theirs (Sec.~\ref{sec:comparison_with_chan}), even this small fraction is enough to offset the depletion seen in \citet{Chan2019}.

\subsubsection{Mass sources in the disk}
\label{sec:mass_sources}

\begin{figure}[!hbt]
  \centering
  \includegraphics[width=\linewidth]{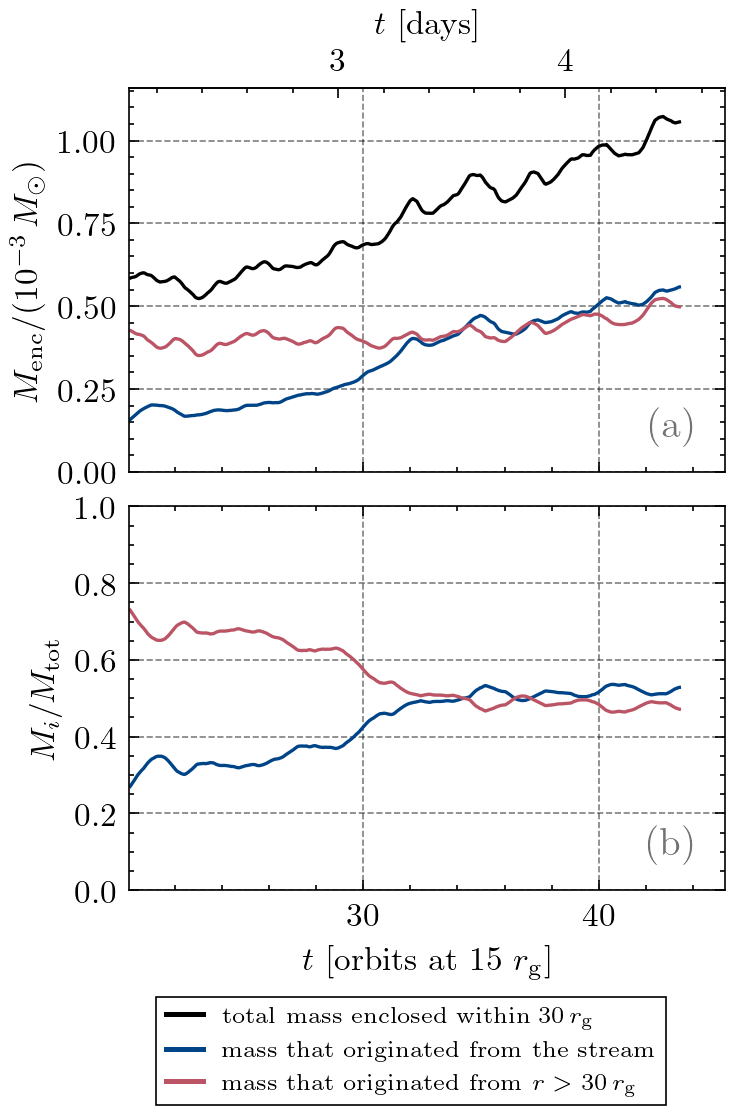}
  \caption{Mass fraction of each component inside the $r<30\,\rg$ sphere as a function of time (in orbits at $15\,\rg$), tagged using passive scalars. Colors are total (black), stream-injected material (blue), and material originating from outside $30\,\rg$ (red). \textbf{(a)} Enclosed mass of different components in units of $10^{-3}\,\Msun$ \textbf{(b)} The different components as a fraction of total mass. At late times, $\simeq 50\%$ of the enclosed mass originated from the stream, and $\simeq 50\%$ originated outside $30\,\rg$.}
  \label{fig:mass_source}
\end{figure}

Fig.~\ref{fig:mass_source} tags the two components and tracks their evolution. Panel (a) shows the enclosed mass of each component, and panel (b) the corresponding mass fractions. 
By the end of the run ($t\sim43$ orbits), the enclosed mass within $30\rg$ is composed of $\sim 50\%$ stream material and $\sim 50\%$ material that was outside $30\,\rg$.
The total enclosed mass itself also grows steadily over this period (panel a), and the persistent rise of the $r>30\,\rg$ contribution shows that mass from well outside the $15\,\rg$ impact radius continues to be advected inward long after the initial collision. We return to this as effect~(3) in Section~\ref{sec:results_flux}.

\subsubsection{Tilted, reorienting inner flow}
\label{sec:disk_tilt}

\begin{figure*}[!hbt]
  \centering
  \includegraphics[width=0.75 \textwidth]{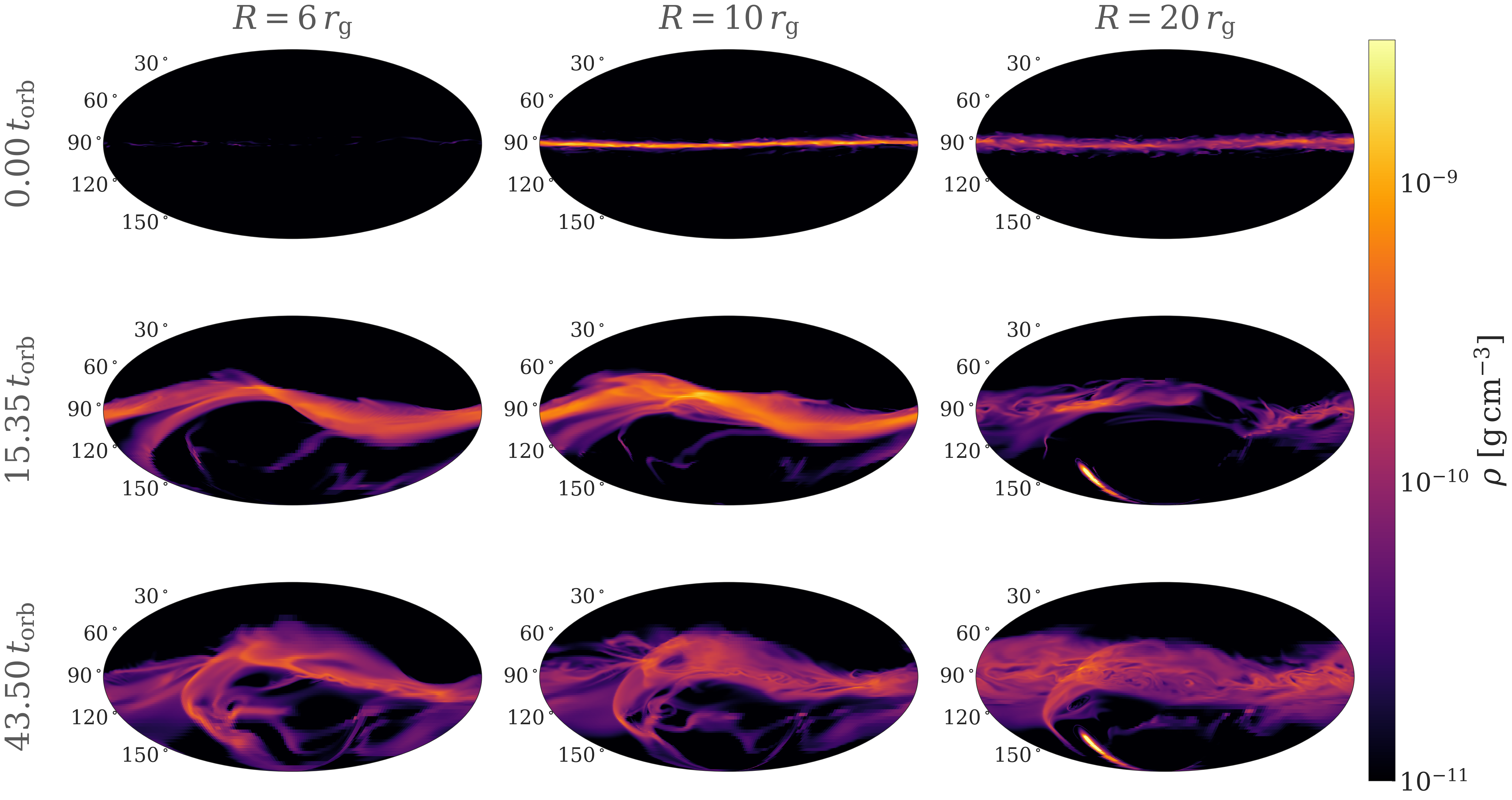}
  \caption{Density as a function of $\theta$ and $\phi$ on three spherical shells,  $r=6$, $10$, and $20\,\rg$ (left to right), shown at three epochs (top to bottom).
   At $t=0$ the gas lies in a thin, axisymmetric equatorial layer that is nearly absent at $r=6\,\rg$ and becomes denser toward $r=20\,\rg$. After stream injection, a vertically oscillating stream wraps in $\phi$ and produces a flow tilted with respect to the equatorial plane that persists to late times; the stream is most coherent at small $r$ and becomes broader, more turbulent, and clumpy at $r=20\,\rg$.}
  \label{fig:rho_thetaphi}
\end{figure*}

Fig.~\ref{fig:rho_thetaphi} maps the gas density in the $\theta$-$\phi$ plane at radii $r=6$, $10$, and $20\,\rg$. Prior to stream injection ($t=0$), the pre-existing disk appears as a thin, axisymmetric equatorial band with nearly $\phi$-independent density. The band is almost absent at $r \leq 6\,\rg$,
the plunging region at and within the ISCO of the pseudo-Newtonian potential,
and becomes progressively thicker and denser toward $r=20\,\rg$.
After the debris stream is introduced, the disk becomes tilted with respect to its initial midplane and appears in these images as a band with a crudely sinusoidal and non-axisymmetric wiggle.

\begin{figure}[!hbt]
  \centering
  \includegraphics[width=\linewidth]{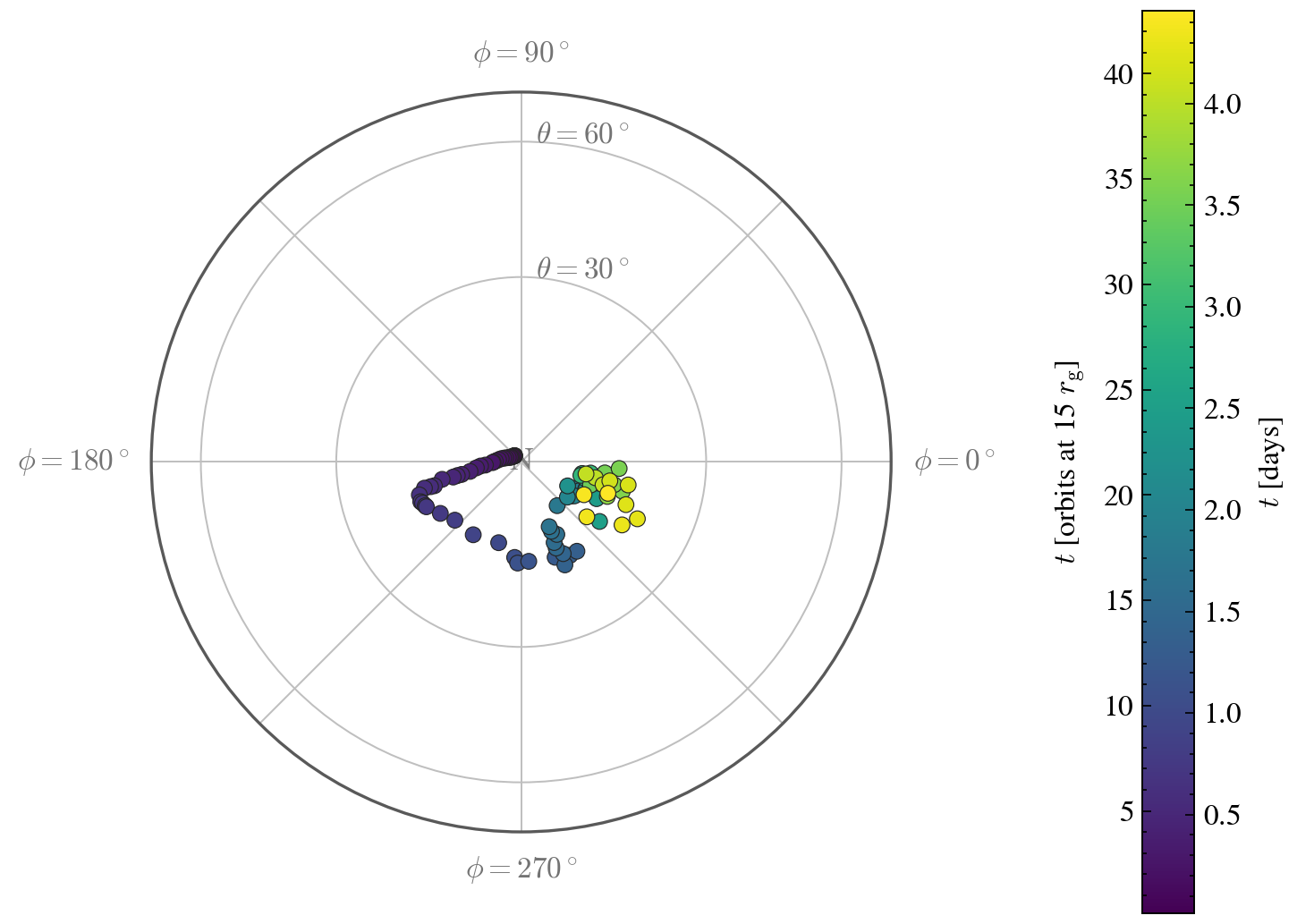}
  \caption{Direction of the total angular momentum of all gas with $r<14\,\rg$ and $\theta\in[\pi/4,3\pi/4]$, shown on a polar projection. The radial coordinate is the tilt angle from the initial disk axis, and the azimuthal coordinate is the orientation of the projected $\vec{L}$; color indicates time in orbits at $15\,\rg$ (outer scale in days). The tilt grows to $30{-}35^\circ$ within $\simeq 15$ orbits, after which the azimuth wanders without a clear period.}
  \label{fig:angular_momentum}
\end{figure}

Fig.~\ref{fig:angular_momentum} shows the direction of the total angular momentum of all gas with $r<14\,\rg$ and $\theta\in[\pi/4,3\pi/4]$ on a polar projection with radius equal to the tilt from the initial disk axis and a rotation angle equal to the azimuth, color-coded by time. The tilt grows to $30{-}35^\circ$ within $\simeq 15$ orbits ($\simeq 1.5$~d). The azimuth sweeps from $\phi\simeq 180^\circ$ through $270^\circ$ to $\phi\simeq 0{-}45^\circ$, then 
wanders randomly in the range $\phi \simeq 340^\circ \pm 20^\circ$.
The driver for this tilt is most plausibly 
momentum transfer from the debris stream to the inner disk.

This tilted flow could generate
disk variability on the same timescale, particularly the wiggling of the inner radiation-flux direction (as seen in Fig.~\ref{fig:mag_and_rad}). In addition, aspect-dependent signatures (obscuration of the corona, or the orientation of an emerging jet) could track the reorientation of the inner flow.

\subsubsection{Preferred inflow trajectory}

\begin{figure*}[!hbt]
  \centering
  \includegraphics[width= \linewidth]{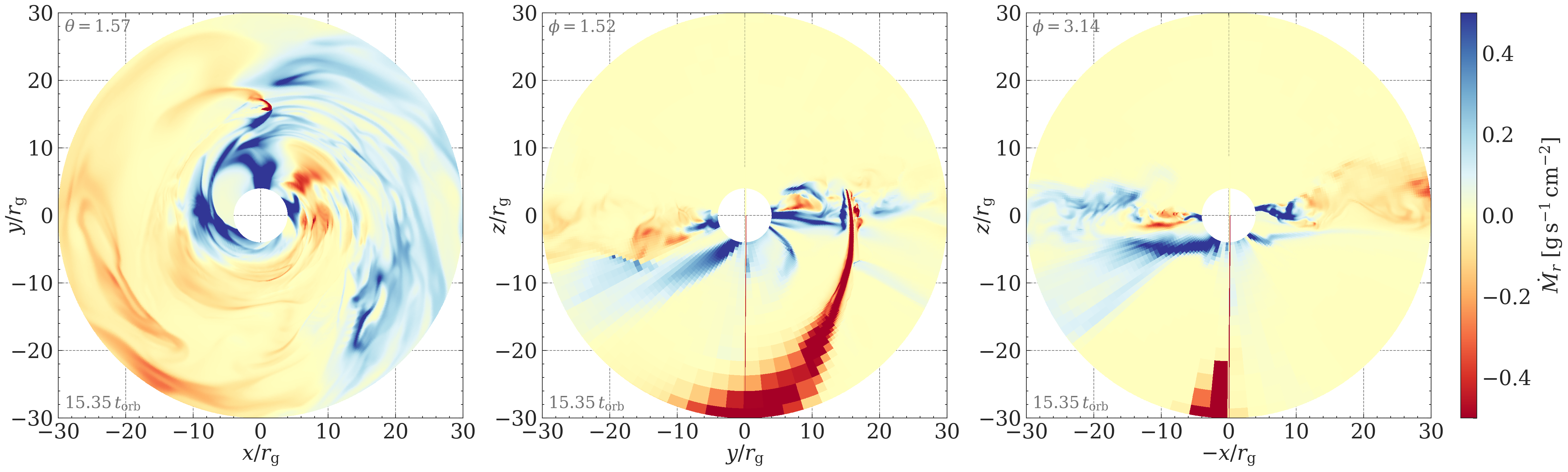}
  \caption{Radial mass flux $-\rho v_r$ at $t=15.35\,\torb$. Left: midplane slice ($\theta=\pi/2$). Center: meridional slice at $\phi=1.52$. Right: meridional slice at $\phi=\pi$. Blue (positive) denotes inflow; red (negative) denotes outflow. Inflow concentrates in a few spiral lanes covering a limited fraction of the azimuth, as in \citet{Chan2019}.}
  \label{fig:m_r}
\end{figure*}

Fig.~\ref{fig:m_r} maps radial accretion rate in $\mathrm{g\,s^{-1}\,cm^{-2}}$ at $t=15.35\,\torb$; positive (blue) values correspond to inflow. In the midplane (left panel), inflow is concentrated in a few broad spiral lanes that cover only a limited fraction of the azimuth and are interleaved with outflow regions. This morphology matches that seen by \citet{Chan2019}.

\begin{figure}[!hbt]
  \centering
  \includegraphics[width= \linewidth]{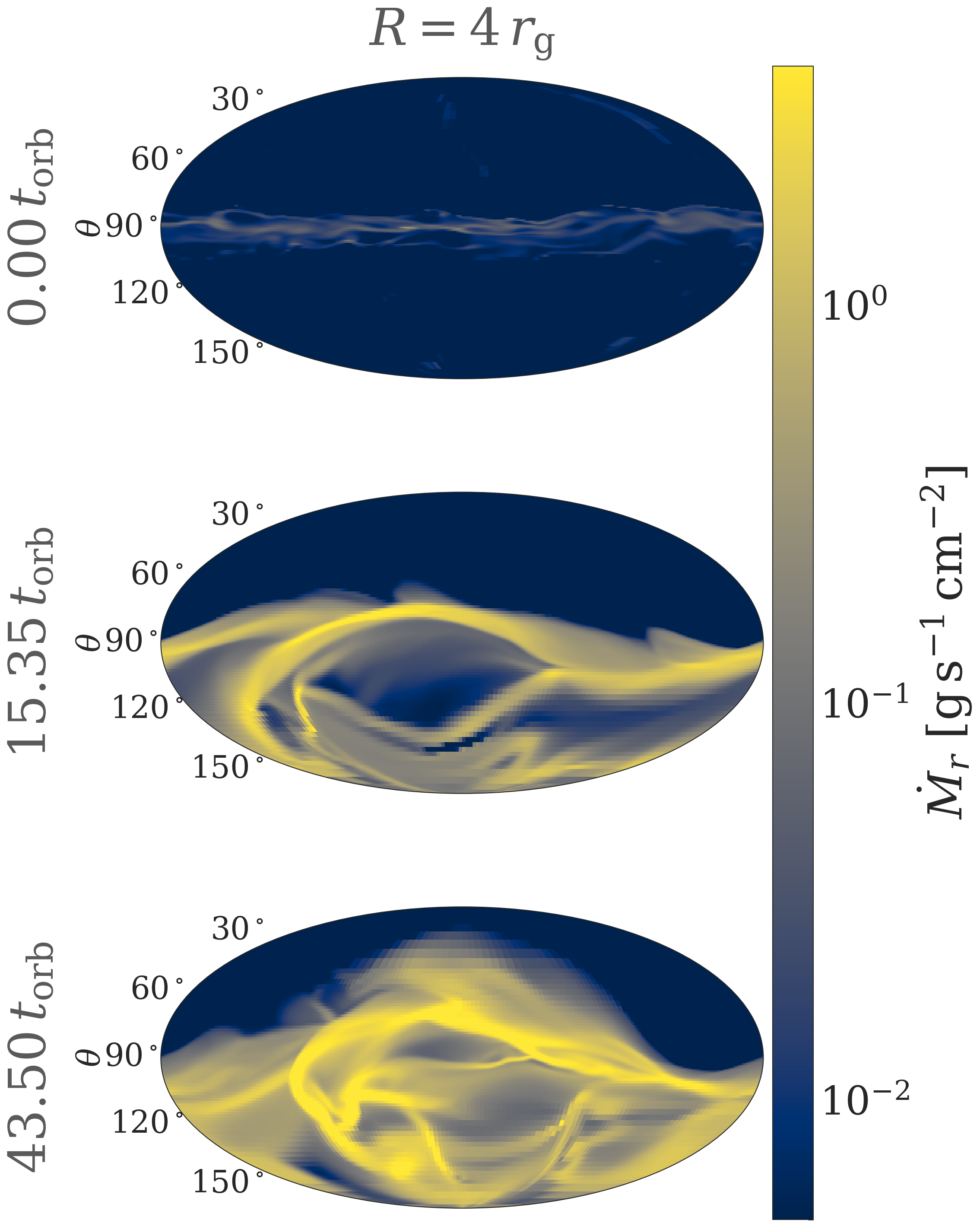}
  \caption{Radial mass flux $-\rho v_r$ on the inner cutout $r=4\,\rg$, shown in the $\theta$--$\phi$ plane at $t=0$, $15.35$, and $43.50\,\torb$.  At $t=0$, the inflow is a thin equatorial belt; after stream injection, the high-flux band follows the tilted midplane (cf.\ Figures~\ref{fig:rho_thetaphi} and~\ref{fig:angular_momentum}), later broadening into filamentary lanes that still avoid the equatorial plane. }  \label{fig:mdot_r_thetaphi}
\end{figure}

To further assess the impact of disk tilt on the inflow trajectory, we show in Fig.~\ref{fig:mdot_r_thetaphi} the inflow on the inner cutout, in units of $\text{g} \, \text{s}^{-1} \, \text{cm}^{-2}$. Prior to stream injection, the accretion through $r=4\,\rg$ is confined to a thin belt at the equatorial midplane, $\theta=\pi/2$. After impact the high-$\dot{M}_r$ band warps into a sinusoidal track in the $\theta$--$\phi$ plane, displaced by $\sim 25^\circ{-}45^\circ$ from the equator by $t=15.35\,\torb$, closely following the tilted midplane of Figures~\ref{fig:rho_thetaphi} and~\ref{fig:angular_momentum}. 
By $t=43.50\,\torb$ the band is broader and more filamentary, spanning much of $|\theta-\pi/2|\lesssim 60^\circ$, indicating that all mass at large tilt angles, and not only the tilted disk, is contributing to the rising accretion rate. This mass could have originated from both the stream and disk matter that was shocked, heated, and lifted to these trajectories. Despite this, most of the mass inflow still takes place through the band occupied by the tilted disk.

\subsubsection{Dynamical response summary}

In short, $\dot{M}$ rises from $0.07\,\dot{M}_{\rm Edd}$ to $\simeq 12\,\dot{M}_{\rm Edd}$ in $4.4$~d while the enclosed mass within $30\,\rg$ grows by a factor $\simeq 5$. 
Roughly $\sim50\%$ of that late-time mass
is disk matter that was outside $30\,\rg$ a few orbits earlier, and the rest is stream mass.
In addition, the inner flow tilts by $30{-}35^\circ$ with a wandering azimuth. The disk puffs up to a thick, lopsided, vertically extended body. 

\subsection{Radiation}
\label{sec:results_lightcurve}

\subsubsection{Bolometric luminosity}

Fig.~\ref{fig:sim_time_series}, panel (d) shows the bolometric luminosity in units of $L_{\rm Edd}$, measured by integrating the vertical radiation flux over the sky-plane footprint of the scattering photosphere (Figs.~\ref{fig:rho_slices} and~\ref{fig:far_rho_slices}), i.e. the luminosity seen by an observer along the $z$ line of sight in each hemisphere. After the stream is introduced, the north-hemisphere luminosity rises from $\simeq 0.02\,L_{\rm Edd}$ to $\sim L_{\rm Edd}$ over the run. The south side peaks near $\simeq 0.1\,L_{\rm Edd}$ within a few orbits and then falls to $\simeq 0.01\,L_{\rm Edd}$: the optically thick outgoing stream obscures thermal photons from escaping and carries them with it. As the stream moves away from the disk, it expands, draining the photon energy by adiabatic expansion work.
The escaping thermal output is therefore dominated by the northern hemisphere and remains $\sim L_{\rm Edd}$.

Over the same interval $\dot{M}$ grows by more than two orders of magnitude relative to its initial value while $L$ stays near Eddington, so the radiative efficiency $\eta \equiv L / \dot{M} c^2$ falls steadily: with $\dot{M}_{\rm Edd}\equiv 10\,L_{\rm Edd}/c^2$, late-time values $L\sim L_{\rm Edd}$ and $\dot{M}\simeq 12\,\dot{M}_{\rm Edd}$ imply $\eta\simeq 0.01$, an order of magnitude below the fiducial thin-disk value $0.1$. This is the trapping-dominated regime anticipated by \citet{Chan2019}, who argued that when the inflow time is shorter than the photon diffusion time, much of the dissipated energy is advected with the gas rather than radiated. In fact, Eddington luminosity is a general result valid for any optically thick system powered by the conversion of gravitational energy into radiation (e.g., accretion). We can check whether this optical thickness criterion holds at the impact radius. Equating the diffusion time, $\sim\tau H/c$, with the inflow time, $\sim r/v_r$, defines a trapping radius $r_{\rm tr} \simeq 2\,(\dot M c^{2}/L_{\rm Edd})\,(H/r)\,\rg \approx 240\,(H/r)\,\rg$ for $\dot M \simeq 12\,\dot M_{\rm Edd}$. For any $H/r \gtrsim 0.1$, this exceeds the impact radius, so the collision region lies inside the trapping zone, and most of the heat dissipated there is carried inward before it can escape.

\subsubsection{Radiation flux and radiation temperature}

\begin{figure*}[!hbt]
  \centering
  \includegraphics[width=\linewidth]{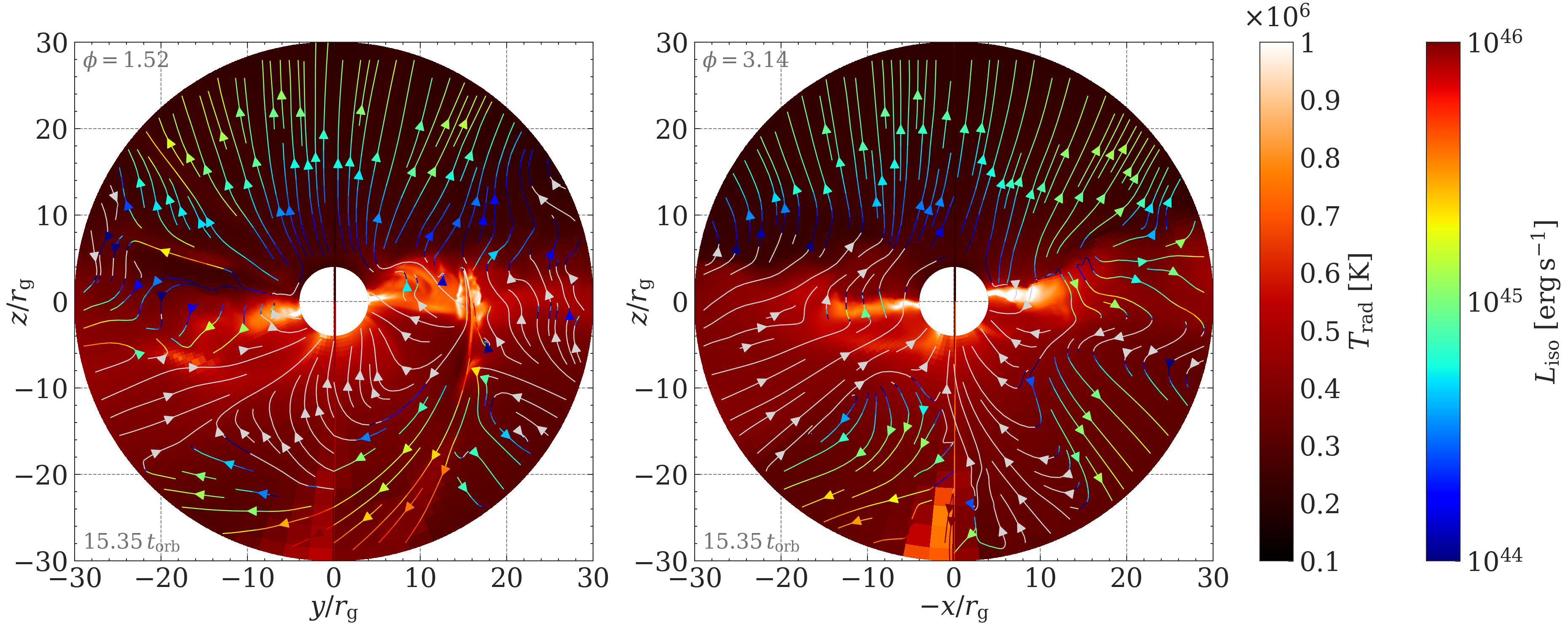}
  \includegraphics[width=\linewidth]{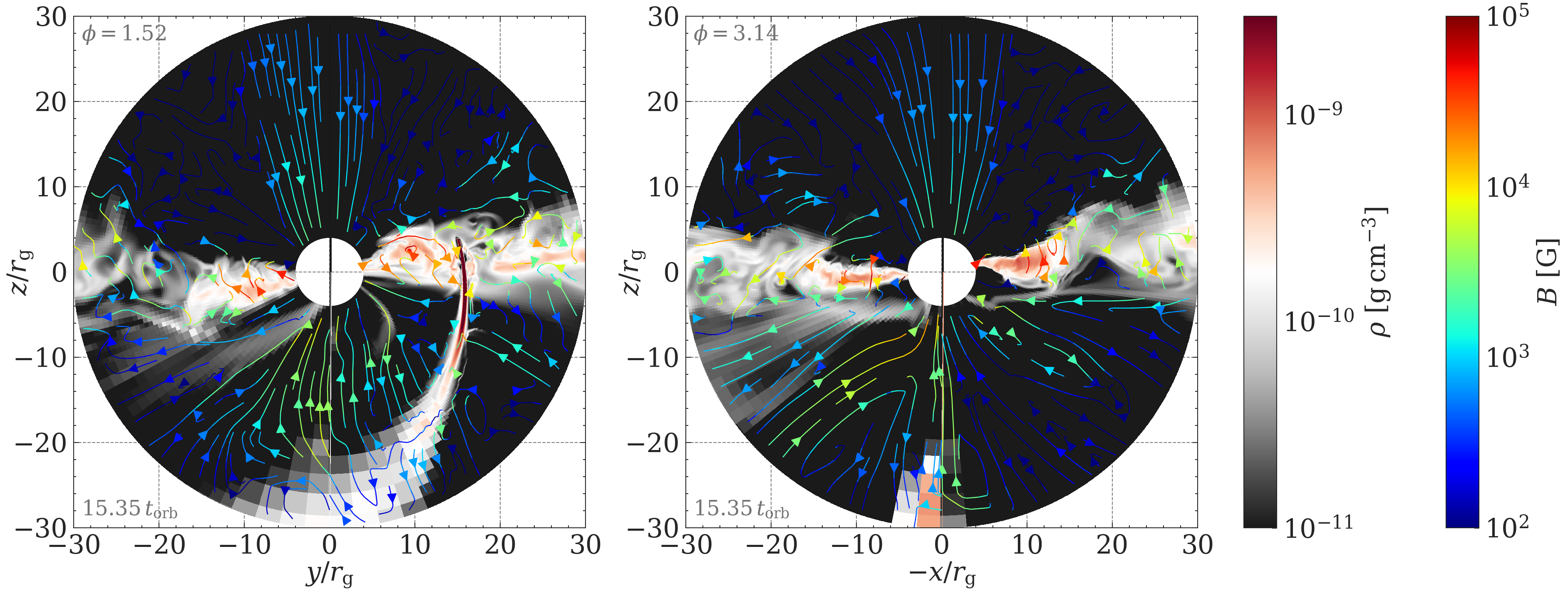}
  \caption{Radiation and magnetic fields on two meridional slices at $t=15.35\,\torb$. \textbf{(Top)} 
  Radiation temperature $T_{\rm rad}=(E_{\rm rad}/a)^{1/4}$ (background) and radiation flux streamlines, colored by the isotropic-equivalent luminosity $L_{\rm iso}=4\pi r^{2}(\mathbf{F}_{\rm rad}\cdot\hat{r})$, ranging from $\simeq 10^{44}$ to $\simeq 10^{46}\,\ergs$, at $t=15.35\,\torb$. Inward-going streamlines are colored gray, since $L_{\rm iso}$ is negative for them.  Near the midplane $\tau\gg 1$, so $\mathbf{F}_{\rm rad}$ is parallel to the flow. In the northern hemisphere, the photosphere is at a few $\rg$ above the disk. Above it, the flux turns outward and escapes. However, the photosphere in the southern hemisphere lies beyond the range of this plot; hence, some of the flux is directed inwards, and even the flux lines directed outwards are trapped. \textbf{(Bottom)} Density (background) with magnetic-field streamlines. Away from the midplane, the field structure near the inner boundary is nearly monopolar, with field lines pointing toward the black hole. This coherent structure could create strong jets. Flux in the southern hemisphere is  $\times$few larger than in the northern one, so jets, if produced, would be $\times$10 stronger in the south.}
  \label{fig:mag_and_rad}
\end{figure*}

Figure~\ref{fig:mag_and_rad} (top) displays the radiation temperature $T_{\rm rad}=(E_{\rm rad}/a)^{1/4}$ overlaid with radiation flux streamlines $\mathbf{F}_{\rm rad}$ at $t=15.35\,\torb$. The innermost disk ($\lesssim 10\,\rg$) near the tilted midplane is the hottest, reaching $T_{\rm rad} \gtrsim 10^6$~K. Across most of the southern hemisphere, out to $r \lesssim 30\,\rg$, the gas maintains $T_{\rm rad} \simeq 5 \times 10^{5}$~K. In the northern hemisphere, while comparable temperatures ($T_{\rm rad} \sim 5 \times 10^5$~K) persist out to $\simeq 5\,\rg$ above the midplane, farther into the northern hemisphere the temperature drops sharply to $T_{\rm rad} \lesssim 10^5$~K.

The streamlines are colored by the isotropic-equivalent luminosity $L_{\rm iso}=4\pi r^{2}(\mathbf{F}_{\rm rad}\cdot\hat{r})$. 
Its absolute value spans the range $\simeq 10^{44}$ to $\simeq 10^{46}\,\ergs$, and is largest along the outgoing stream in the southern hemisphere, where it approaches $\sim 10^{46}\,\ergs$. It is important to note that the entirety of the southern hemisphere inside $30\rg$ is inside the photosphere, and there are significant portions of that region where the flux is directed {\it inward.} 
Inside the photosphere, the optical depth is large, so radiation is advected with the gas and $\mathbf{F}_{\rm rad}$ runs parallel to the flow; only above the photosphere does the flux turn and escape.
The flux is strong in the outgoing stream, but those photons do not immediately escape: they remain trapped and are carried south with the debris.  Similarly, the regions where the nominal $L_{\rm iso}$ is negative have inward-directed fluid velocity (see Fig.~\ref{fig:rho_slices}). 

\begin{figure*}[!hbt]
  \centering
  \includegraphics[width=\linewidth]{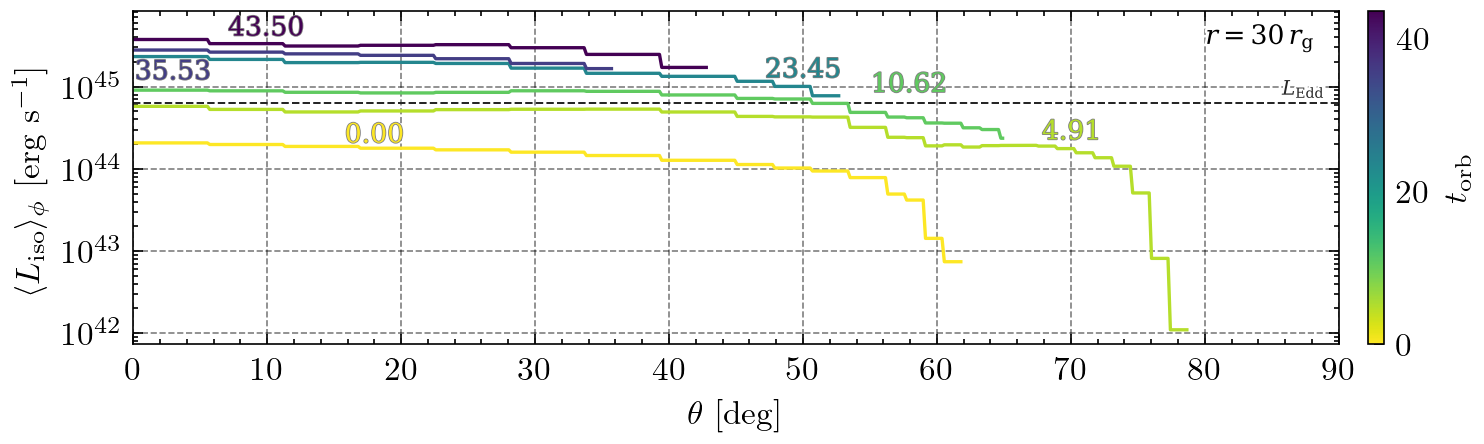}
  \caption{Azimuthally averaged isotropic-equivalent luminosity $\langle L_{\rm iso}\rangle_\phi$ defined by $L_{\rm iso}=4\pi r^{2}(\mathbf{F}_{\rm rad}\cdot\hat{r})$, at $r=30\,\rg$ versus polar angle $\theta$, measured from the north pole. Color and labels give time in orbits at $15\,\rg$. Each curve stays high over a limited range of $\theta$ and then falls sharply at the point where the sphere $r=30\,\rg$ intersects  
  the photosphere. That cutoff is near $60^\circ$ before impact and retreats to $\sim 40^\circ$--$45^\circ$ by $t\simeq 43\,\torb$ as the photosphere extends vertically (Fig.~\ref{fig:far_rho_slices}). On the axis, $\langle L_{\rm iso}\rangle_\phi$ grows from $\simeq 2\cdot 10^{44}\,\ergs$ ($\simeq 0.3\,L_{\rm Edd}$) to $\simeq 3\cdot 10^{45}\,\ergs$ ($\simeq 5\,L_{\rm Edd}$).}
  \label{fig:L_iso_theta}
\end{figure*}

Elsewhere, especially in most of the northern hemisphere, $L_{\rm iso}$ is consistently positive, but lower in magnitude, typically $\sim 10^{44}$--$10^{45}\,\ergs$, and is at the higher end of that range for most of the northern hemisphere outside the photosphere (shown at this time in the middle row of Fig.~\ref{fig:rho_slices}). Fig.~\ref{fig:L_iso_theta} shows, for a number of other times during the simulation, the azimuthally averaged $L_{\rm iso}(\theta)$ at $r=30\,\rg$ (outside the photosphere). We see that each curve is nearly flat in polar angle and then drops at the angle where this radius passes inside the photosphere. After impact, the cutoff first moves out to $\theta\sim 80^\circ$, then retreats as the photosphere extends vertically (Fig.~\ref{fig:far_rho_slices}), so that by $t\simeq 43\,\torb$ the escaping flux is confined to a cone of opening angle $\sim 40^\circ$--$45^\circ$. Across that cone $L_{\rm iso}$ is roughly flat at $\simeq 5\,L_{\rm Edd}$. The polar value rises from $\simeq 0.3\,L_{\rm Edd}$ before impact to $\simeq 5\,L_{\rm Edd}$ at the end of the run.

Because the overall optical depth of the southern hemisphere is greater than in the north, the photon-bright hemisphere is the north (Fig.~\ref{fig:sim_time_series}, panel~d).  
On the other hand, the south is the jet-bright side, carrying the larger polar-cap magnetic flux (Section~\ref{sec:results_flux}). 

\subsubsection{Photosphere and emitted photon energies}

The photospheres marked in Figs.~\ref{fig:rho_slices} and~\ref{fig:far_rho_slices} provide a direct estimate of the emitting surface. After stream injection, the south photosphere follows the outgoing stream to $|z|$ of several hundred $\rg$ (Fig.~\ref{fig:far_rho_slices}): that surface is larger than the northern photosphere, but it is the obscured, thermally dim side.

\begin{figure*}[!hbt]
  \centering
  \includegraphics[width=0.85\linewidth]{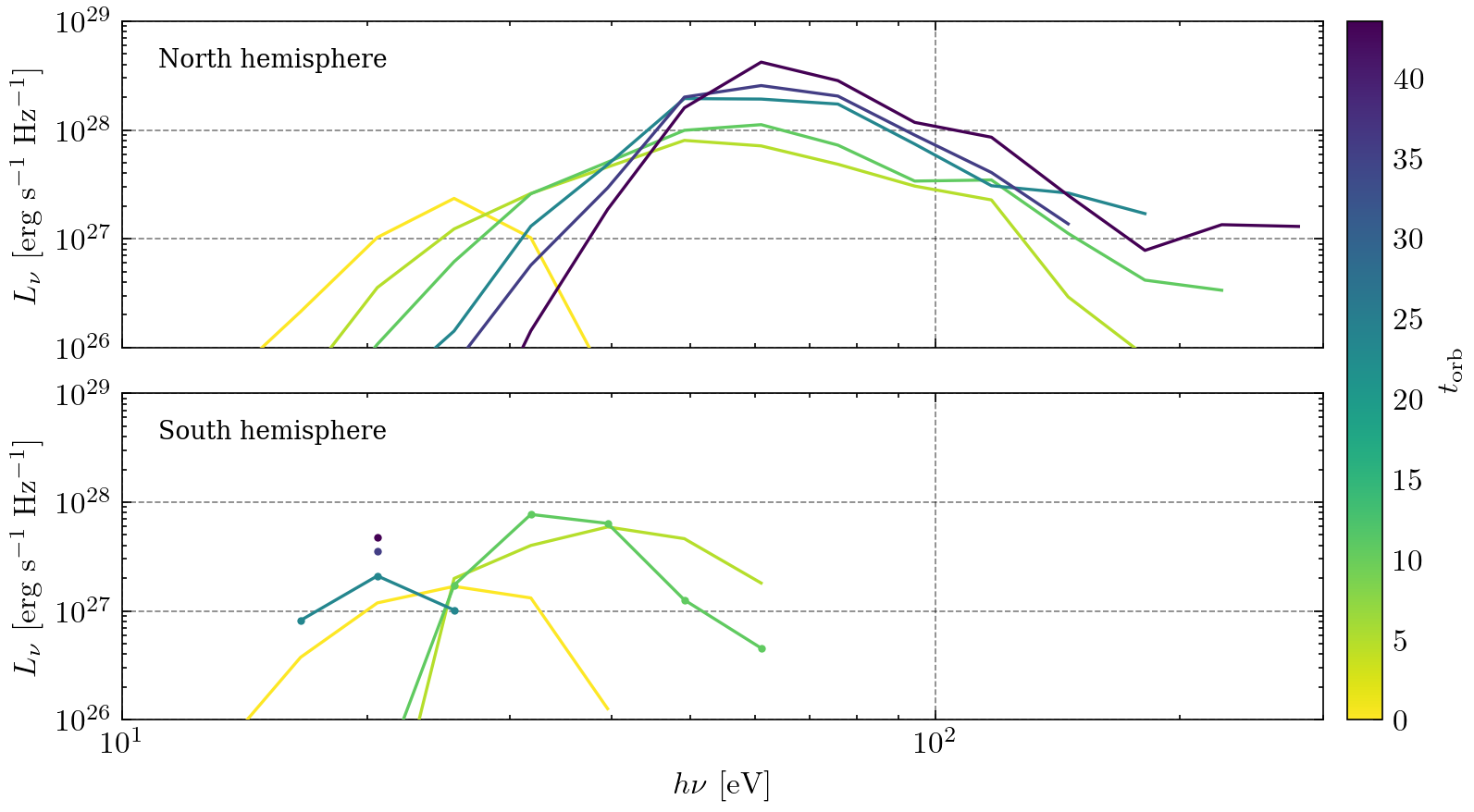}
  \caption{Spectral luminosity $L_\nu$ reconstructed from the thermalization-photosphere flux, shown separately for the north and south hemispheres and colored by time. Each photosphere patch is assigned an effective temperature from its vertical flux and contributes at the corresponding Wien peak, so $\int L_\nu\,\mathrm{d}\nu$ equals the bolometric $L$ of Fig.~\ref{fig:sim_time_series}, panel~(d). The emission is in the EUV, peaking between $\sim 10$ and $\sim 100$~eV; the thermally photon-bright north side hardens over time while the south remains softer. South curves are sparser because that photosphere is more extended and not always fully identified.}
  \label{fig:Lnu}
\end{figure*}

Fig.~\ref{fig:Lnu} reconstructs a spectral luminosity from the thermalization-photosphere flux: each sky-plane patch is assigned an effective temperature from its vertical flux and contributes at the corresponding Wien peak, so $\int L_\nu\,\mathrm{d}\nu$ equals the bolometric $L$. The resulting $L_\nu$ peaks between $\sim 10$ and $\sim 100$~eV, in the EUV. The thermally photon-bright north side hardens over time (peak shifting from $\sim 20{-}30$~eV to $\sim 50{-}80$~eV) while the south remains softer ($\sim 20{-}50$~eV).

\subsubsection{Radiation summary}

The escaping luminosity, measured on the scattering photosphere, is dominated by the thermally photon-bright north hemisphere and approaches $\sim L_{\rm Edd}$ while $\dot{M}$ climbs to $\simeq 12\,\dot{M}_{\rm Edd}$.  The radiative output stays near Eddington even though the inflow is highly super-Eddington, corresponding to a radiative efficiency $\eta\simeq 0.01$ under our $\dot{M}_{\rm Edd}=10\,L_{\rm Edd}/c^2$ convention. The south is thermally dim because the stream obscures photons and carries them with it, and is the jet-bright side. $L_\nu$ reconstructed from the thermalization-photosphere flux peaks at $10{-}100$~eV, in the EUV. 

\subsection{Magnetic-flux transport toward the black hole}
\label{sec:results_flux}

Can the stream-disk interaction deliver enough horizon-threading flux, fast enough, to power a transient jet? As we argued in \citet{Tripto2026}, a viable mechanism must (i)~supply magnetic flux of order $\Phi\sim 10^{29}\,\mathrm{G\,cm^{2}}$ when $M_{\rm BH}\sim 10^{6}\,\Msun$ in order to generate a jet luminosity $\sim 10^{44}\,\ergs$, and (ii)~act within a time $\sim t_0$, typically several weeks, of the onset of mass return.

\begin{figure}[!hbt]
  \centering
  \includegraphics[width= \linewidth]{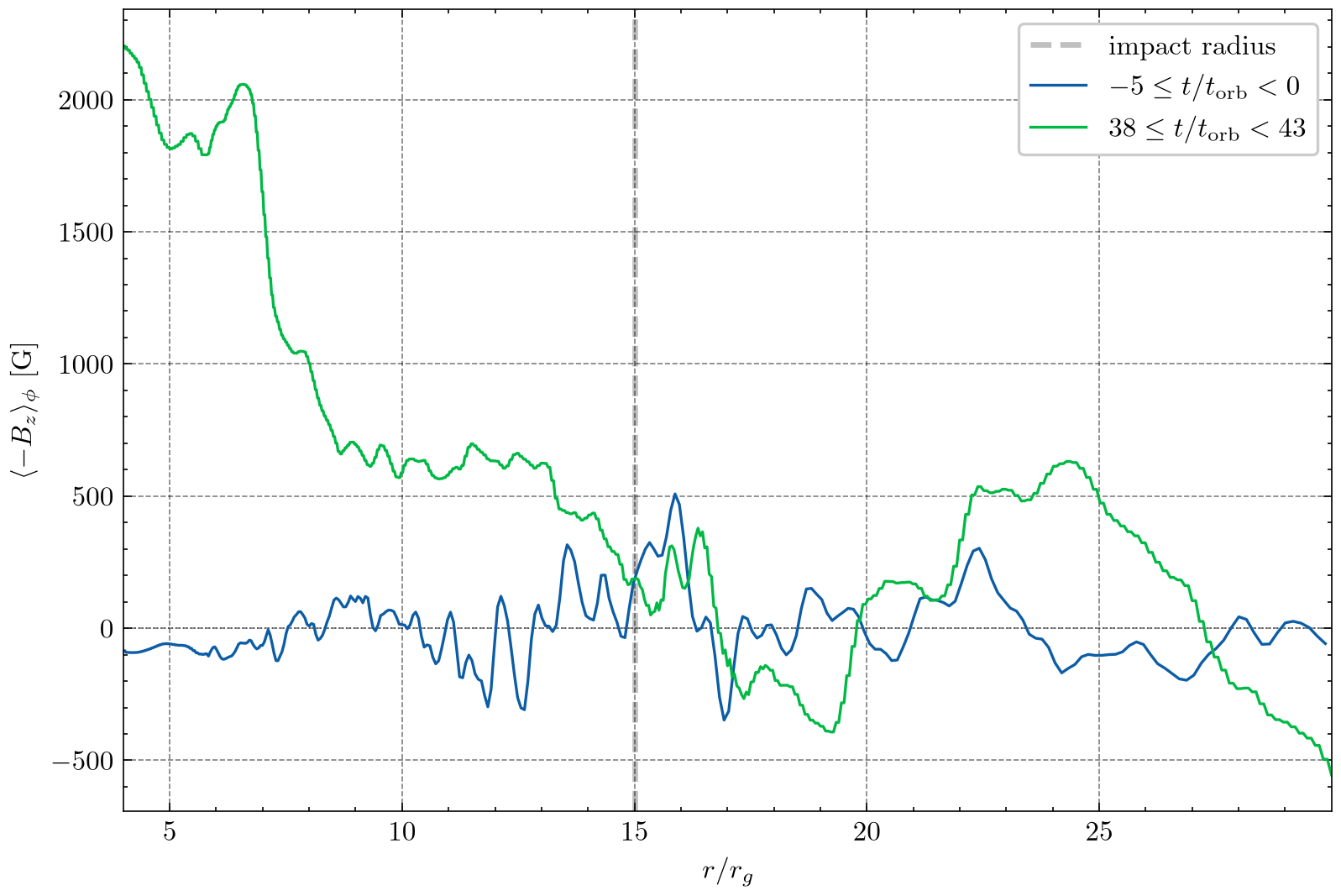}
  \caption{$\phi$-averaged midplane $-B_z$ as a function of radius, time averaged over a 5-orbit window before stream injection ($-5$–$0\,\torb$) and at the end of the simulation ($38$–$43\,\torb$). Before impact, the profile alternates in sign with $|B_z|\lesssim500$~G. After impact, it is dominated by a strong positive peak of $\simeq2200$~G, decaying to $\simeq600$~G by $10$–$13\,\rg$. The field is substantially amplified and gains a more coherent structure after the impact.}
  \label{fig:B_z}
\end{figure}

The quantity relevant to jet launching is the ordered flux threading the polar region, as seen in Figure~\ref{fig:mag_and_rad} 
(bottom).  Because these images show that the flux is very coherent within the black hole's polar caps, we 
define $\Phi$ as the flux on the inner cut-out at angles $< \pi/4$ from the polar axis.
The AGN0.07 run, initialized with a quadrupolar field, wholly lacks such order at the midplane (Fig.~\ref{fig:B_z} demonstrates the disorder on the midplane at the moment the debris is first injected).
Figure~\ref{fig:B_z} also shows that the $\phi$-averaged midplane $-B_z$ changes from a sign-alternating profile with $|B_z|\lesssim 500$~G before impact to a single-signed inner profile peaking at $\simeq 2200$~G afterward. Thus, our simulation shows that the debris impact also organizes the vertical field through the disk midplane, as discussed in \citet{Tripto2026}.

Fig.~\ref{fig:sim_time_series}, panel (c) shows the polar-cap flux as a function of time. Prior to stream injection, the flux on the two poles is comparable, as expected, and flat at $\Phi\simeq 6\cdot 10^{27}\,\mathrm{G\,cm^{2}}$ over the preceding $13$ orbits. Within $1{-}3$ orbits ($0.1{-}0.3$~d), the mean flux jumps by a factor $\simeq 4$ to $\simeq 2.5\cdot 10^{28}\,\mathrm{G\,cm^{2}}$, and the fluxes on the two poles begin to depart from each other. The southern hemisphere flux then grows slowly, eventually reaching 
$\simeq 2\cdot 10^{29}\,\mathrm{G\,cm^{2}}$ by orbit~$30$ ($3$~d) and then saturating; meanwhile, the north-hemisphere flux also rises slowly until $\simeq 20$~orbits and thereafter fluctuates in the range $\sim 2\times 10^{28}$--$10^{29}$~G~cm$^2$.

Three mechanisms appear to contribute to that growth.

\textit{Effect~(1): collapse of pre-existing flux interior to the impact radius.} The rapid inflow discussed in Section~\ref{sec:results_mdot_mass} carries the disk's associated field inward. This alone can account for the initial factor of $\simeq 4$ increase, to a mean flux $\simeq 2\cdot 10^{28}\,\mathrm{G\,cm^{2}}$. 
The first few orbits are exactly the time during which the material originally in the inner disk is swept toward the horizon. The pre-impact flux in the disk, estimated by the cone integral described below, is $\simeq 10^{28}\,\mathrm{G\,cm^{2}}$, matching this initial rise. It is the sole effect we considered in our baseline analytic estimate \citep{Tripto2026}.

\begin{figure}[!tb]
  \centering
  \includegraphics[width=0.9\linewidth]{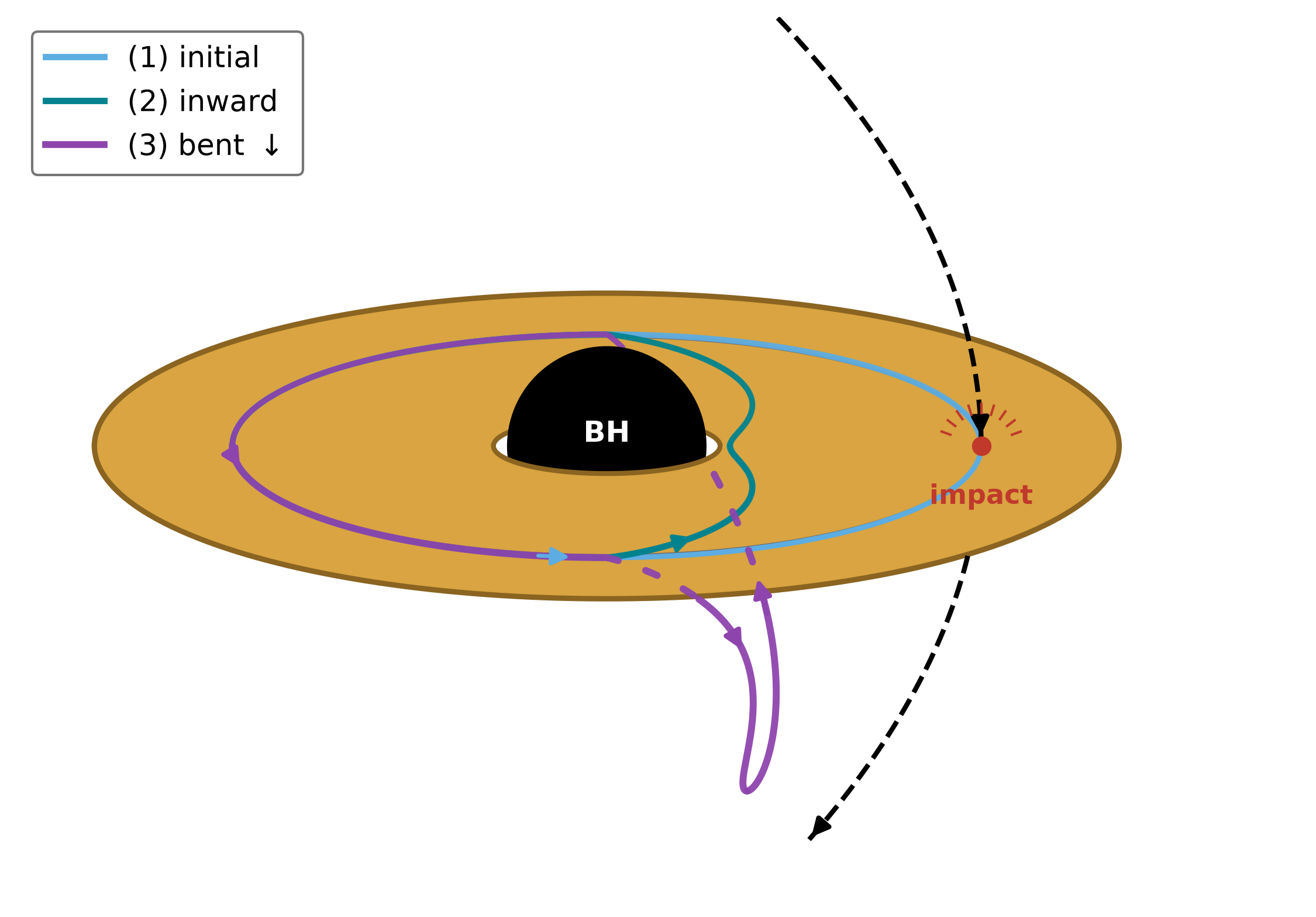}
  \caption{Schematic of the proposed toroidal-to-poloidal conversion mechanism. A representative field loop initially occupies a narrow radial range in the orbital plane (1, light blue). The debris stream impact advects its near side inward toward the black hole, stretching the loop over a wide range of $r$ (2, teal). The near side is then displaced out of the orbital plane by $\sim90^\circ$ while the far side remains in the plane (3, purple), so the loop comes to thread the polar cone on the side of the vertical displacement, here downward, toward the lower cone.}
  \label{fig:toroidal_field_organization}
\end{figure}

\textit{Effect~(2): conversion of toroidal field to poloidal field.}
Fig.~\ref{fig:toroidal_field_organization} illustrates this mechanism, in which the poloidal flux on the inner cut-out grows when toroidal fieldlines are bent into poloidal fieldlines while leaving the poloidal flux in the remainder of the domain untouched.
The initial (pre-disk) quadrupolar field is wound by orbital shear into a toroidal field of coherent sign over an extended radial range (loop 1); because such a loop closes on itself without leaving the disk plane, 
this winding does not add poloidal flux. The impact of the debris stream converts part of this toroidal reservoir into poloidal flux through two geometric deformations of field lines. 
First, matter attached to these field lines moves inward, pulling its portion of the field lines inward toward the black hole and stretching the field lines across a wide range of $r$ (loop 2). 
Next, the debris pushes matter in the inner disk out of the disk plane, in most cases moving it into the southern hemisphere, while matter on the same field lines that remains near its original radius stays approximately in the disk plane (loop 3). 
As a result, the attached field lines are bent by $\sim90^\circ$, and poloidal flux is created in the black hole polar caps. 
The direction of the debris's motion therefore accounts for the observed upper/lower-cone flux asymmetry (Fig.~\ref{fig:sim_time_series}, panel (c)).
This description is qualitatively consistent with the simulation's field morphology (Fig.~\ref{fig:mag_and_rad}, bottom). It also explains why $B_z$ in the inner disk becomes predominantly negative, as seen in Figure~\ref{fig:B_z}.  Figure~\ref{fig:toroidal_field_organization} shows that field on a polar cap with a given sign requires vertical field of the opposite sign somewhere in the disk; Figure~\ref{fig:mag_and_rad} (bottom) shows that the majority of the polar cap flux, the part on the south side, has $B_z > 0$; the mean $B_z$ in the inner disk must therefore be negative.

\textit{Effect~(3): advection from outside the impact radius.} Fig.~\ref{fig:mass_source} shows that $\simeq 50\%$ of the late-time mass inside $30\,\rg$ originated from outside $30\,\rg$, well beyond the $15\,\rg$ impact point. Mass advected from outside the impact point carries additional magnetic flux, further increasing the flux available for advection onto the black hole. We noted this possibility in \citet{Tripto2026} as well, but excluded it from that analysis to obtain a conservative estimate.

Effect~(1) delivers $\simeq 10^{28}\,\mathrm{G\,cm^{2}}$; effects~(2) and~(3) raise the collected flux by about an order of magnitude to $\simeq 2\cdot 10^{29}\,\mathrm{G\,cm^{2}}$. Unfortunately, we cannot distinguish the contributions of these two mechanisms in the simulation.

Adopting a slightly modified version of the conventions of \citet{Tchekhovskoy2011} and \citet{KelleyTchekhovskoy2014} (we discuss these conventions more carefully in \citealt{Tripto2026}), the Blandford-Znajek scaling gives $L_{\rm BZ}=c\xi_{\rm BZ}\Phi^{2}/(2\pi^{2}\rg^{2})$. With $\xi_{\rm BZ}=0.1$, which corresponds to a rapidly spinning black hole,
\begin{equation}
\begin{aligned}
L_{\rm BZ}\simeq 3\cdot 10^{42}\,\ergs\,
&\left(\frac{\xi_{\rm BZ}}{0.1}\right)
\left(\frac{\Phi}{10^{29}\,\mathrm{G\,cm^{2}}}\right)^{2} \\
&\times \left(\frac{M_{\rm BH}}{5\cdot 10^{6}\,\Msun}\right)^{-2}.
\end{aligned}
\end{equation}
The polar flux is not symmetric between the two hemispheres: during the middle of the simulation, the flux threading the southern pole reaches $\sim 2\cdot10^{29}\,\mathrm{G\,cm^{2}}$, raising that side's isotropic-equivalent luminosity to $\sim  10^{43}\,\ergs$. 
These values are obtained with a quadrupolar (zero-net-flux in each hemisphere) initial field. Even higher values could be achieved with a more favorable magnetic configuration (e.g., dipolar) initial field whose inner region carries net poloidal flux \citep{BHK2008}.

The magnetic flux on the south polar cap exceeds that in the north by a factor of 3 - 10, corresponding to a Poynting flux ratio $\sim 10 - 100$.
The south is therefore the jet-bright hemisphere, opposite the thermally photon-bright north. There is a hint of decline on the favored (south) side after $\simeq 3$~d that is not seen on the north; confirming this would require a longer run with a time-dependent fallback rate. The selection-effect implication of the hemisphere asymmetry is deferred to Section~\ref{sec:discussion}.

These specific quantitative results can be compared with the approximate predictions of our analytic scheme \citep{Tripto2026}, which used the vertical magnetic flux on the midplane within the impact radius, here $r = 15\rg$.
Because the quadrupolar initial state makes the vertical field on the midplane disordered, we choose to measure the pre-impact flux on a narrow cone around the polar axis. In Figure~\ref{fig:sim_time_series}, we plotted the flux on the cut-out, the end-product, so to speak; here we measure the amount of ``raw materia''.
Integrating $\rg^2\int_0^{2\pi} d\phi \, \int_4^{15}\, dr \, B_\theta(\theta=7\pi/16)\,r\sin \theta $
gives an estimate of the flux in the disk $\simeq 10^{28}\,\mathrm{G\,cm^{2}}$ before impact, 
roughly twice the flux on the cut-out at the same time.  This is an order of magnitude smaller than the flux achieved through the action of the debris stream.  In other words, the impact of the tidal debris enlarges the available flux by a factor $\sim 10$ relative to the simple analytic estimate of the available flux.

\subsubsection{Magnetic flux summary}

The simulation reaches 
$\Phi\simeq 2\cdot 10^{29}\,\mathrm{G\,cm^{2}}$ in the south after $\sim 30$ orbits at $15\rg$ ($3$ d), which is higher than the pre-existing flux by a factor of 
$\sim 33$. For our black hole mass, $5\cdot 10^{6}\,\Msun$, that flux can lead to a jet luminosity 
$\sim 10^{43}\,\ergs$, which would be sufficient for some jetted TDEs. 
This flux is delivered to the black hole horizon by three mechanisms. Our analytic estimate \citep{Tripto2026} used only one: the accumulation of flux from inside $r_{\rm imp}$; including toroidal-poloidal conversion and outer-radius advection, the flux grows by about an order of magnitude.
The underlying quadrupolar field geometry of our simulation did not yield a consistent vertical-field sign; if it had, for example, a dipolar structure, the flux might have been even greater \citep{BHK2008}.

\section{Discussion}
\label{sec:discussion}

\subsection{Where the Matter Goes; Comparison with Previous Work}

\citet{Chan2019} reported a set of purely hydrodynamic simulations illustrating what would happen when a TDE debris stream strikes an accretion disk.  In addition to omitting both radiation transfer and magnetic fields, these simulations used the $\alpha$-model to determine the initial state of the disk. Replacing that model with the AGN0.07 disk raises $\dot{M}_s/\dot{M}_d$ by about two orders of magnitude (to $\simeq 400$, versus $\le 4$). Even so, our results, which incorporate radiation transfer for gas cooling and radiation forces, as well as magnetic fields, while also using a more physically realistic initial disk, confirm several of the key results of the earlier work.  In only a few local disk orbits, the mass contained in the disk at radii inside $r_{\rm imp}$ is swept rapidly toward the black hole's event horizon at a rate that can be $\sim O(10 \dot{M}_{\rm Edd})$.  Throughout the debris-disk interaction, spiral shocks are driven outward from the impact points, and the gas inflow is concentrated in azimuthally localized channels. 
However, there are several ways that the additional physics we have introduced changes the outcome of the event.

First, the ratio $\dot{M}_s/\dot{M}_d$ is so large that the stream's momentum completely dominates the collision between it and the disk gas.  The result is that the disk gas is strongly deflected, while the stream is deflected only slightly. 

Second, two new mechanisms copiously resupply the disk's inner region with matter.  Magnetic stresses and radiation forces push matter from as far away as $\simeq 2 r_{\rm imp}$ to well inside $r_{\rm imp}$.  In addition, a significant fraction of the mass brought to the disk in the debris stream is diverted so that it, too, finds itself at $r < r_{\rm imp}$ en route to the horizon; because $\dot{M}_s/\dot{M}_d$ is so large, the diverted debris mass is significant relative to the disk mass.  Between the two of these sources, the mass in the inner disk at $\simeq 40$ local orbits after the debris first arrived is $\approx 3 \times$ what it was at the start (as shown in Fig.~\ref{fig:surface_density} (bottom)).

Third, the force of the debris stream striking the disk from the ``north'' pushes the disk gas ``southward'', causing the disk to tilt.  Over time, the irregularity of the mass distribution in the impact region is so great that, as shown in Figure~\ref{fig:rho_thetaphi}, the inner disk is better described as a collection of mutually-tilted streams. 

Fourth, the collision barely changes the stream's momentum, but it does reshape its orbit.
After the impact, about a third of the debris keeps roughly its original energy, nearly two-thirds becomes more tightly bound (a fifth of the debris by more than a factor of two), and only a few percent becomes less bound.
Thereafter, the debris moves outward nearly ballistically beneath the bulk of the disk, turning around at $\sim 50$--$400\,\rg$; the most bound fifth returns within a few orbital periods at $r_{\rm imp} = 15\,\rg$, while the rest 
returns to the disk plane at larger radii ($\gtrsim O(100) \, \rg$) at later times.

We expect that the debris that has spread out will be stopped by the outer disk, and that its fate will resemble that discussed by \citet{Chan2021}. An exact calculation of this phase lies beyond the extent and duration of our simulation.

\subsection{Photon Emission}

The thermal luminosity in both hemispheres first jumps by a factor $\sim 10$ within a few hours of the first interaction, to $\simeq 0.1\,L_{\rm Edd}$, and then grows more slowly, reaching $\sim L_{\rm Edd}$ by $\sim 4$~d.  The luminosity from the northern hemisphere grows more slowly thereafter, reaching $\sim L_{\rm Edd}$ by $\sim 4$~d. Because $L_{\rm Edd}$ is the characteristic luminosity for any gas mass suddenly heated by dissipation of kinetic energy obtained from its gravitational infall and then cooled by photon diffusion regulated by electron scattering within the same mass
\citep{Krolik2010}, the peak luminosity is unlikely to vary substantially for systems with different parameters.  Meanwhile, the luminosity from the southern hemisphere falls, dropping to $\simeq 1/2$ its initial value over that same time period.  As a result, the north side is $\simeq 150\times$ brighter than the south side.  The cause of this sharp contrast is the debris stream matter, just noted, which makes the southern hemisphere considerably optically thicker than the northern.

The spectrum reconstructed from the photosphere peaks in the EUV, at $10{-}100$~eV (Fig.~\ref{fig:Lnu}).  So many astrophysical environments have large opacity to photons in this band that any detection of this continuum is likely to be through reprocessing 
within the galactic nucleus. 
Because we are considering TDEs in AGN, the nearest such reprocessor would be the 
AGN's broad emission line region.  If the greatest emission line response from the broad line gas comes from gas with the optimal ionization parameter (i.e., ratio of ionizing flux to density), the distance to the BLR adjusts $\propto L^{1/2}$.  That this might be so is suggested by the strong $r \propto L^{1/2}$ correlation found when distance is estimated from the variability time lag.  Applying a bolometric correction of 5 to the standard luminosity normalization
of this correlation \citep{Bentz+2013} 
indicates $r \simeq 13 L_{\rm bol,44}^{1/2}$~lt-d.  Strictly speaking, this correlation applies to Balmer lines, and it is possible that the spectrum we expect from these TDEs would more efficiently produce higher-ionization lines, so the lag might be somewhat shorter.  In any event, one might expect a lag of roughly this magnitude, which is only a factor of a few shorter than typical values of the characteristic TDE timescale $t_0$.

There could likewise be a reprocessed signal from absorption in the dusty obscuring torus.  The distance between its inner edge and the black hole should likewise scale $\propto L^{1/2}$ because it is most likely determined by dust sublimation.  Calibrating this relation to NGC~1068 \citep{Jaffe+2022}, we find $r \sim 80 L_{\rm bol,44}^{1/2}$~lt-d.  This reprocessing could therefore create a mid-IR flare echo with a delay of a few months.

Two interesting questions about the detectability of photons from TDEs in AGN remain open.  One is the shape of the lightcurve over timescales of a few $t_0$; our simulation was far too short to shed any light on this question. The other is that both our simulation and that of \citet{Chan2021} suggest that a significant fraction of the debris mass falls back to the accretion disk at radii $\sim 10 - 100 \times r_{\rm imp}$, or $\sim 10^2 - 10^3 \rg$.
Some of this mass might therefore land on the region where the undisturbed AGN disk radiates its optical/UV continuum, the photons we most easily observe.  \citet{Chan2021} argued that the energy released when debris falls back on the disk would emerge as hard X-rays; it is possible that there may also be an augmentation of the optical/UV luminosity.

\subsection{Magnetic Flux and the Implied Jet Luminosity}

As estimated in Sec.~\ref{sec:results_flux}, the generic simulation we conducted brought $\sim 10^{29}$~G~cm$^2$ of magnetic flux to the black hole in a time only a fraction of $t_0$. On the basis of this simulation, which was essentially Newtonian, we cannot speak to the luminosity of any resulting jet.  However, if the black hole spins rapidly enough to provide a reasonable jet efficiency ($\xi_{\rm BZ} \sim 0.1$), the expected jet power would be $\sim 10^{43}$~erg~s$^{-1}$, comparable to several of the observed jetted TDEs.  
Recall that \citet{Tripto2026} argued that the presence of a magnetized accretion disk is the most favorable mechanism for providing the magnetic field needed to power jets in TDEs.

Like the thermal photon luminosity, the magnetic flux on the black hole polar caps is strongly asymmetric, but in the opposite sense: it is larger on the ``south'' pole than on the ``north'' by a factor of $\sim 3$--$10$.  With the expected jet luminosity scaling as $\Phi^2$, the luminosity asymmetry should be $\sim 10$--$100$.  This effect is also due to symmetry breaking arising from the debris stream's direction of travel, a feature that distinguishes TDEs in AGN from ordinary AGN.

Very much unlike the thermal photon luminosity, the jet luminosity could be sensitive to a great many parameters.  It increases rapidly with black hole spin \citep{McK2005,HK2006,Tchekh2010,Tchekhovskoy2011}. It should also vary with the magnetic field geometry: as shown by \citet{BHK2008}, dipolar fields support jets better than the quadrupolar field of our simulation.  Inclination of the stellar orbit relative to the disk material's orbit (presumably aligned with the black hole spin) should also matter.   The magnetic flux in the disk can be expected to rise with accretion rate and depend upon $M_{\rm BH}$, while the interaction with the debris stream will also change with $\Ms$.  Consequently, observed jet luminosities are likely to deviate considerably from this single example. 

The magnetic flux delivered to the black hole increases with the magnetic flux in the pre-existing disk inside $r_{\rm imp}$, but is not strictly limited to it because the action of the TDE itself can raise the flux given to the black hole by an order of magnitude or more: in the fraction of $t_0$ covered in our simulation, flux originally located as far away as $\simeq 2 r_{\rm imp}$ can be brought in; in addition, toroidal field can be rotated so as to be poloidal when carried to the horizon. 
However, the fact that the flux in the disk is limited does imply that the magnetic field strength on the black hole horizon does not necessarily rise to the limiting level associated with magnetically arrested disk (MAD) states.   In our example, the normalized flux $\phi \equiv \Phi/\sqrt{\dot{M} \rg^2 c} \propto \Phi (\dot{M}/\dot{M}_{\rm Edd})^{-1/2} \MBH^{-3/2}$ remains of order $\sim 0.1$. It changes by only a factor of a few while the flux grows by a factor $\sim 33$, because $\dot{M}$ grows at the same time. This value of $\phi$ is more than two orders of magnitude below the prerequisite for MAD.
Nonetheless, the associated jet luminosity can be quite substantial. The limited variation of $\phi$ also implies both that the magnetic pressure just outside the black hole is regulated to have somewhat less than the ram pressure of infalling matter and that the associated jet's efficiency with respect to the accretion rate stays of the same order. 

\subsection{Observed Population}

Nearly all the power in thermal photons emitted at the photosphere of the inner disk region is radiated at energies in the range $\sim 10 - 100$~eV.  Through the combined action of the many ionization edges found in that band and dust absorption, it is very difficult for such photons to travel more than a short distance through either the galactic nucleus in which they are generated or the ISM of our own galaxy.  Consequently, even when our view is toward the bright side (the ``north'' side) of the AGN disk, this spectrum will only rarely be visible.   In contrast, the spectra of jets are much broader, and large parts of their radiated power are often in the range $10 - 100$~keV, in which only the lowest end may have significant absorption opacity.  This fact makes the jet radiation much more readily observable, especially when we are on the ``south'' side of the disk.

On the other hand, as already remarked, the absorption of thermal photons that renders them unobservable directly can create new photons that are easier to detect, either atomic emission lines or mid-IR thermal radiation from dust grains.  Combining these several facts suggests that in observed samples, jetted TDEs will in most instances be seen from the ``south'' side of the AGN disk, and only through their jet radiation and possibly emission lines or mid-IR. The reprocessed signals are not subject to the same close-in obscuration as the thermal photon continuum, so they should be observable from either the north or the south side. In the mid-IR, the flare should be delayed by several months relative to the jet radiation flare.
We caution, however, that our simulation was designed to mimic the period of peak mass-return rate in the TDE.  It therefore cannot predict any specific time-dependence for either the thermal photon luminosity or the jet luminosity.

Finally, because the fraction of galaxies possessing AGN is $\sim 10^{-1}$ and the fraction of AGN having strong jets is also $\sim 10^{-1}$, the fraction of all TDEs resulting in strong jets should be $\sim 10^{-2}$, consistent with the empirical estimate \citep{Andreoni+2022}.

\section{Summary}

As first shown by \citet{Chan2019}, the debris dynamics of a TDE in an AGN differ drastically from those of an ordinary TDE; as we show here, these altered dynamics produce somewhat greater observed luminosities, radiate much more asymmetrically, and have quite different spectra than ordinary TDEs.
They can also supply the magnetic flux needed to launch a jet.

The stream--disk collision excites spiral shocks that drive rapid, super-Eddington accretion. At first, the collision carves a midplane cavity. This is followed by mass build-up in the innermost disk, reaching eventually a steady plateau of the surface density.   Ultimately, the inflow into the black hole is about two-thirds captured stream material and one-third native disk gas fed from the outer region. Furthermore, interaction with the incoming stream rapidly tilts the inner disk out of its original plane on dynamical timescales, inducing irregular azimuthal precession.

The source of the photons is also entirely different from that in ordinary TDEs. In the ordinary case, the optical/UV light draws its energy from shocks at $\sim 10^3\,\rg$ and is radiated from this shock region. When the TDE occurs in an AGN, the energy comes from shocks within a few tens of $\rg$ and is radiated from a photosphere of complex geometry, whose distance from the black hole ranges from $10$ -- $20\rg$ in the north hemisphere to several hundred $\rg$ in the south; the photosphere at $10$ -- $20\rg$ is much brighter thermally. The escaping luminosity in the brighter (northern) direction reaches $\sim L_{\rm Edd}$ or more, and $L_\nu$ peaks at $10{-}100$~eV, in the EUV.

We have shown that the polar cap magnetic flux can grow by a factor $\sim 33$ and then saturate during a small fraction of a characteristic mass-return time.
Three effects contribute: the collapse of pre-existing flux interior to the impact radius, the organization of toroidal field into poloidal flux, and advection of flux from outside the impact radius. Our earlier estimate \citep{Tripto2026} retained only the first; the present run strengthens that estimate by about an order of magnitude. The resulting jet luminosity, $\sim 10^{43}\,\ergs$, is consistent with some jetted TDEs.

Both the thermal emission and the jet are highly anisotropic, but they favor opposite hemispheres. The north (the side from which the stream arrives) is thermally photon-bright and dominates the escaping luminosity. The south (the direction of stream motion) is thermally dimmer because the stream obscures photons, 
but its polar-cap flux exceeds the north's by a factor of $\sim 3$--$10$ near the flux peak, leading to a factor of $\sim 10$--$100$ in Poynting flux, making the south the jet-bright side.

\section*{Acknowledgments} 
We thank Yan-Fei Jiang for providing the input data for our simulation and for extensive consultations about Athena++. We also thank him, Chi-Ho Chan, and Scott Noble for helpful discussions. This work was supported by the Simons Foundation SCEECS collaboration (MP-SCMPS-00001470) (NT,TP), by an ERC advanced grant MultiJets (NT,TP), and by NSF Grant 2110339 and NASA TCAN grant 80NSSC24K0100 (JHK).

\FloatBarrier
\clearpage
\appendix
\numberwithin{equation}{section}

\section{Unit conversion}
\label{app:unitconverstion}

We initialize the ambient disk using a snapshot from the ``AGN0.07'' global RMHD simulation of \citet{Jiang2019}, which was carried out for a black hole mass
$M_{\rm BH,0}=5\cdot10^8\,\Msun$.
In this work we require a similar disk solution around
$M_{\rm BH}=5\cdot10^6\,\Msun$.

In Newtonian hydrodynamics one can often rescale mass/length/time units with multiple free parameters, but in radiation-MHD, the presence of universal dimensional constants and fixed microphysics reduces or eliminates this freedom.
More generally, the number of free rescaling parameters is $N_f=\max(0,3-N_{\rm udc})$, where $N_{\rm udc}$ counts independent universal dimensional constants appearing in the equations and in the prescribed microphysics/boundary conditions \citep{Granot2012}.
Therefore, in RMHD with gravity there is no non-trivial unit rescaling that preserves the full set of equations; one must instead choose a mapping that keeps a selected set of parameters fixed.
We keep (i) radii in units of $\rg$, (ii) the Eddington ratios $\dot M/\dot M_{\rm Edd}$ and $L/L_{\rm Edd}$, and (iii) the RMHD dimensionless parameters used by Athena++ equations ($\mathbb{C}\equiv c/v_*$ and $\mathbb{P}\equiv aT_*^4/(\rho_* v_*^2)$) invariant. A similar procedure, followed for much the same reasons, is described in \citet{Zhu+2026}.

Let $M\equiv M_{\rm BH}$ denote the target mass and $M_0\equiv M_{\rm BH,0}$ the original one. For a generic physical quantity $X$ given in cgs units, we denote its code value equivalent as $\bar{X}$, and the conversion factor from code units to physical units as $X_* = X/\bar{X}$.
We choose the characteristic units $(r_*,v_*,\rho_*)$ to scale as
\begin{equation}
\begin{aligned}
  r_* &\propto M,\\
  v_* &\propto M^0,\\
  \rho_* &\propto M^{-1}.
\end{aligned}
\end{equation}
which implies
\begin{equation}
\begin{aligned}
  t_*\sim r_*/v_* &\propto M,\\
  P_*\sim \rho_* v_*^2 &\propto M^{-1},\\
  T_* \propto P_*^{1/4} &\propto M^{-1/4},\\
  B_*\sim \sqrt{\rho_*}\,v_* &\propto M^{-1/2}.
\end{aligned}
\end{equation}
With these choices, the luminosity unit scales as $L_*\sim T_*^4 r_*^2\propto M$, so $L/L_{\rm Edd}$ stays constant by construction, and the electron-scattering optical depth scale $\kappa_{\rm es}\rho_* r_*$ is also invariant.

The mass ratio is $M/M_0=10^{-2}$. The implied conversion factors are therefore
\begin{equation}
\begin{aligned}
  r_* &= 10^{-2}\, r_{*,0}, &
  t_* &= 10^{-2}\, t_{*,0}, &
  v_* &= v_{*,0},\\
  \rho_* &= 10^{2}\,\rho_{*,0}, &
  P_* &= 10^{2}\,P_{*,0}, &
  T_* &= 10^{1/2}\,T_{*,0},\\
  B_* &= 10\,B_{*,0}. & & &
\end{aligned}
\end{equation}

This mapping preserves the electron-scattering opacity, but not the absorption opacity, which depends explicitly on $\rho$ and $T$. The AGN0.07 disk was deliberately placed near the iron opacity bump ($T\approx1.8\cdot10^5\,$K; \citealt{Jiang+2016}), whereas after rescaling ($T_*=10^{1/2}\,T_{*,0}$$=(M/M_0)^{-1/4}\,T_{*,0}$) the gas lies at $T\approx6\cdot10^5\,$K, where the Rosseland-mean absorption opacity is substantially lower. The rescaled disk is therefore not an exact thermal equilibrium of our equations and evolves on its own even without stream injection. This drift is slow compared with the stream-driven evolution: over the $\approx13$ orbits preceding injection (Fig.~\ref{fig:sim_time_series}, $t<0$), the accretion rate, magnetic flux, luminosity, and mass within $30\,\rg$ remain nearly constant, and the mass within $12\,\rg$ grows by $\approx60\%$, roughly an order of magnitude slower than its growth after injection.

\section{Stream Injection Scheme}
\label{app:stream}

Following the general prescription of \citet{Chan2019}, the TDE debris is introduced through a source term in the hydrodynamic equations. The injection region is described in parabolic coordinates $(\ell,m,n)$, in which the spherical radius is $r = (\ell+m)/2$ and $n$ is an azimuthal angle. This stream-aligned frame is rotated into the simulation frame by the angles $(\theta_{\rm pitch},\theta_{\rm roll},\theta_{\rm yaw})$.

The injected density has a Gaussian profile centered at $(\ell_0,m_0,n_0)$,
\begin{equation}
\rho_{\rm s} = \rho_{\rm c} \exp\!\left[-\left(\frac{\ell-\ell_0}{\Delta\ell}\right)^{2} -\left(\frac{m-m_0}{\Delta m}\right)^{2} -\left(\frac{n-n_0}{\Delta n}\right)^{2}\right],
\end{equation}
where $(\Delta\ell,\Delta m,\Delta n)$ set the size of the injection region and $\rho_{\rm c}$ is normalized to the prescribed mass current $\dot M_{\rm s}$. The corresponding physical widths at the injection point are $\simeq 0.1\,\rg$ and $\simeq 0.3\,\rg$ transverse to the stream, i.e.\ $\simeq \Rsun$ and $\simeq 3\Rsun$ ($\sim 1.2$--$4\Rs$).

The injected velocity is directed along the tangent of a Newtonian parabolic orbit, with magnitude $v_{\rm s} = (\eta_{\rm v}GM_{\rm BH}/r)^{1/2}$; $\eta_{\rm v} = 2$ is the marginally bound value in a Newtonian potential. At the center of the injection region, this places the stream at the pericenter of its orbit, $r_p \simeq 15\,\rg$, with specific angular momentum $\simeq 5.7\,\rg c$ and inclination $i \simeq 86^\circ$ to the disk. In the Paczy\'nski--Wiita potential, $\eta_{\rm v} = 2.1$ leaves the stream mildly bound, $E \simeq -6\cdot10^{-3}c^2$, compared with $|\Delta E| \sim 4\cdot10^{-4}c^2$ for debris near peak return. This affects only gas that crosses the disk without strong interaction, since the collision changes the specific energy by $\sim GM_{\rm BH}/r_p \sim 0.07\,c^2$; we find no stream material returning from the ballistic apocenter ($\sim 160\,\rg$) during the run.

The stream pressure is set by an isothermal sound speed, $P=\rho c_{\rm s}^2$, so the source term injects mass, momentum, and gas internal energy. No magnetic field is injected with the debris, and the source term does not directly add radiation energy. Changes in the magnetic flux near the black hole therefore come from the response of the pre-existing AGN field to the impact, and radiation associated with the stream appears only after the injected gas shocks, heats, and couples to the radiation field. The parameter values are listed in Table~\ref{tab:stream-parameters}.

\begin{table}
    \centering
    \begin{tabular}{lc}
    \hline
    Parameter & Value \\
    \hline
    $(\theta_{\rm pitch},\theta_{\rm roll},\theta_{\rm yaw})$ & $(0,\,-1.9,\,1.5)$ \\
    $(\ell_0,m_0,n_0)$ & $(7.8,\,23,\,1.7)$ \\
    $(\Delta\ell,\Delta m,\Delta n)$ & $(0.1,\,0.05,\,0.025)$ \\
    Injection point $(r,\theta,\phi)$ & $(15.4,\,1.3,\,1.5)$ \\
    $\eta_{\rm v}$ & $2.1$ \\
    $c_{\rm s}$ & $2.6\cdot 10^{6}\,\mathrm{cm\,s^{-1}}$ \\
    $\dot{M}_{\rm s}$ & $8.3\cdot 10^{25}\,\mathrm{g\,s^{-1}}$ \\
    \hline
    \multicolumn{2}{l}{Derived orbit at the injection center:} \\
    Pericenter $r_p$ & $15.6$ \\
    Inclination to the disk $i$ & $86^\circ$ \\
    \hline
    \end{tabular}
    \caption{Fiducial stream parameters. Lengths are in $\rg$; angles, including $n_0$ and $\Delta n$, are in radians.}
    \label{tab:stream-parameters}
\end{table}

\FloatBarrier
\clearpage

\end{document}